\documentclass{emulateapj}
\usepackage{epstopdf}
\usepackage{graphics}
\usepackage{natbib}
\newcommand{\beq}	{\begin{equation}}
\newcommand{\eeq}	{\end{equation}}
\newcommand{\beqa}{\begin{eqnarray}}
\newcommand{\eeqa}{\end{eqnarray}}

\def\simlt{\lower.5ex\hbox{$\; \buildrel < \over \sim \;$}}
\def\simgt{\lower.5ex\hbox{$\; \buildrel > \over \sim \;$}}

\font\tenbi=cmmib10 
\newfam\bifam  \textfont\bifam=\tenbi
\font\tenbr=cmbx10
\newfam\brfam  \textfont\brfam=\tenbr

\font\squinttenbi=cmbx10 at 9pt
\scriptfont\brfam=\squinttenbi

\def\vecnabla{
              \setbox1=\hbox{$\bigtriangledown$}
                           \raise.45ex\hbox{$\bigtriangledown$\hskip-.97\wd1
                           $\bigtriangledown$\hskip-.97\wd1
                           $\bigtriangledown$\hskip-.97\wd1}
                           \raise.47ex\hbox{$\bigtriangledown$}}

\def\rsun{\ifmmode {\rm R}_{\mathord\odot}\else $R_{\mathord\odot}$\fi}
\def\msun{\ifmmode {\rm M}_{\mathord\odot}\else $M_{\mathord\odot}$\fi}
\def\lsun{\ifmmode {\rm L}_{\mathord\odot}\else $L_{\mathord\odot}$\fi}

\def\tmb{\ifmmode {T_{\rm mb}^{13}(x,y,v)}\else $T_{\rm mb}^{13}(x,y,v)$\fi}

\begin{document}

\title{Modeling the Atomic-to-Molecular Transition and Chemical Distributions of Turbulent Star-Forming Clouds}

\author{Stella S.~R.~Offner\footnote{Hubble Fellow}}
\affil{Department of Astronomy, Yale University, New Haven, CT 06511}
\email{stella.offner@yale.edu }

\author{Thomas G.~Bisbas}
\affil{Department of Physics and Astronomy, University College London, Gower Street, London WC1E 6B}

\author{Serena Viti}
\affil{Department of Physics and Astronomy, University College London, Gower Street, London WC1E 6B}

\author{Thomas A.~Bell}
\affil{ Centro de Astrobiolog\'ia (CSIC-INTA), Carretera de Ajalvir, km 4, 28850 Madrid, Spain}

\begin{abstract}
We use {\sc 3d-pdr}, a three-dimensional astrochemistry code for modeling photodissociation regions (PDRs), to post-process hydrodynamic simulations of turbulent, star-forming clouds. We focus on the transition from atomic to molecular gas, with specific attention to the formation and distribution of H, C$^+$, C, H$_2$ and CO. First, we demonstrate that the details of the cloud chemistry and our conclusions are insensitive to the simulation spatial resolution, to the resolution at the cloud edge, and to the ray angular resolution. We then investigate the effect of geometry and simulation parameters on chemical abundances and find weak dependence on cloud morphology as dictated by gravity and turbulent Mach number. For a uniform external radiation field, we find similar distributions to those derived using a one-dimensional PDR code. However, we demonstrate that a three-dimensional treatment is necessary for a spatially varying external field, and we caution against using one-dimensional treatments for non-symmetric problems. We compare our results with the work of Glover et al.~(2010),~who self-consistently followed the time evolution of molecule formation in hydrodynamic simulations using a reduced chemical network. In general, we find good agreement with this in situ approach for C and CO abundances.  However, the temperature and H$_2$ abundances are discrepant in the boundary regions (Av $\le$ 5), which is due to the different number of rays used by the two approaches.
%SSRO , which is likely due to the more approximate PDR treatment implemented by Glover et al.
\end{abstract}

\section{Introduction}

In the local universe, stars appear to form exclusively in cold, dense clouds of predominately molecular gas \citep{mckee07}. Understanding the evolution of these molecular clouds (MCs) and the formation of stars within them is a fundamental problem in astrophysics that is hampered by distance, projection effects, and the high optical depth in these regions. Probing the mass and velocity distributions of the gas is further complicated by the fact that the most abundant molecule, H$_2$, lacks a dipole moment. The next most abundant molecule, CO, which is commonly used to probe the cold molecular gas distribution in lieu of H$_2$, has a typical average abundance of about one per $10^4$ H$_2$ molecules in the Milky Way. In addition, the relationship between CO abundance and total gas mass is a complicated one that depends upon metallicity, the three-dimensional radiation field, the abundances of other molecules, and dust chemistry \citep{bell06, glover11, shetty11}. Accurately modeling the formation of H$_2$ and the relative abundances of homologous molecules such as CO requires following complex chemical reaction networks that encompass hundreds of species and thousands of reactions.

Traditionally, the computational expense of evolving large chemical networks limited astrochemical investigations to simple one-dimensional hydrodynamic models (e.g., \citealt{bergin04}) or to post-processing (e.g., \citealt{levrier12}). However, in recent years ``reduced'' chemical networks have been adopted to investigate chemistry concurrently with three-dimensional hydrodynamics \citep{nelson97, nelson99, pavlovski02, pavlovski06, glover07a, glover07b, glover10}. Such methods have the advantage of being able to follow the temperature evolution of the gas due to UV heating and atomic and molecular cooling, which in principle influences the gas dynamics since shock jump conditions depend upon the local temperature. 
%Both post-processing and insitu astrochemistry calculations have limitations. In post-processing, many more chemical reactions and and gas-grain physics can be included. 
Nonetheless, the expense of following the molecular evolution in situ necessitates various simplifications, including neglect of dust physics and coarse treatment of the radiation field. 

%Note: Simon mentioned 2012a, MNRAS, 421, but that doesn't have self-gravity
Thus far, turbulent cloud calculations including simplified chemistry have also focused on larger cloud complexes and generally neglected the self-gravity of the gas (see \citealt{glover12} as an exception including gravity). Neglecting gravity obviates the need for considerable additional resolution which would otherwise be required to resolve collapsing gas \citep{truelove97}. In addition, without forming embedded sources to provide additional radiation (e.g., \citealt{Offner09a, krumholz07}), heating depends only on the external cloud environment, leading to simpler radiative conditions. The gas temperature range induced by a standard external interstellar radiation field is generally limited ($\lesssim$100K) and deviates from 10 K mainly at low Av.
%Removed, ``boundary'' is ill-defined, cf Simon
%, namely, near the cloud boundaries.  

Despite such simplifications, the astrochemistry under investigation is rich and not well understood. For example, cloud boundary regions are especially interesting because this is where gas transitions from being ionized and atomic to predominantly molecular. These low-Av transitions areas are by definition PDRs, where FUV photons dominate the energy balance and gas chemistry. PDRs are ubiquitous in the interstellar medium and are the source of most of the infrared radiation in galaxies. The recent development of {\sc 3d-pdr} (\citealt{bisbas12}, hereafter B12), which is the first dedicated PDR code able to treat arbitrary three-dimensional density distribution, now allows the accurate study of these regions in more complex structures.

We dedicate this paper to three main goals. First, we compare 3D and 1D treatments of a complex PDR region in order to evaluate the impact of dimensionality on chemical results. Thus, we extend the work of B12, who demonstrated the importance of higher dimensional treatment in accurately modeling simple 3D problems, to consider complex, turbulent gas distributions. Second, we use self-gravitating, hydrodynamic simulations of molecular clouds with different Mach numbers to evaluate the importance of underlying physical parameters on chemical abundances and distributions. Finally, we explore the differences between two astrochemistry approaches by considering results obtained via post-processing using {\sc 3d-pdr} and results obtained from a chemical network calculation preformed ``in situ'' (e.g., \citealt{glover10}).

The paper is organized as follows. In section \ref{methods} we describe the {\sc 3d-pdr} methodology and our hydrodynamic numerical simulations. In section \ref{validation} we validate our choice of spatial resolution by presenting convergence studies of grid-sampling in the cloud interior and at the cloud boundaries.  We present our results in section \ref{results}, including a comparison to \citet{glover10} and discussions of chemical dependence on domain dimensionality, external radiation field, and cloud physical parameters. Section \ref{conclusions} contains a discussion of future work and conclusions.

\begin{figure*}
\epsscale{1.0}
\plotone{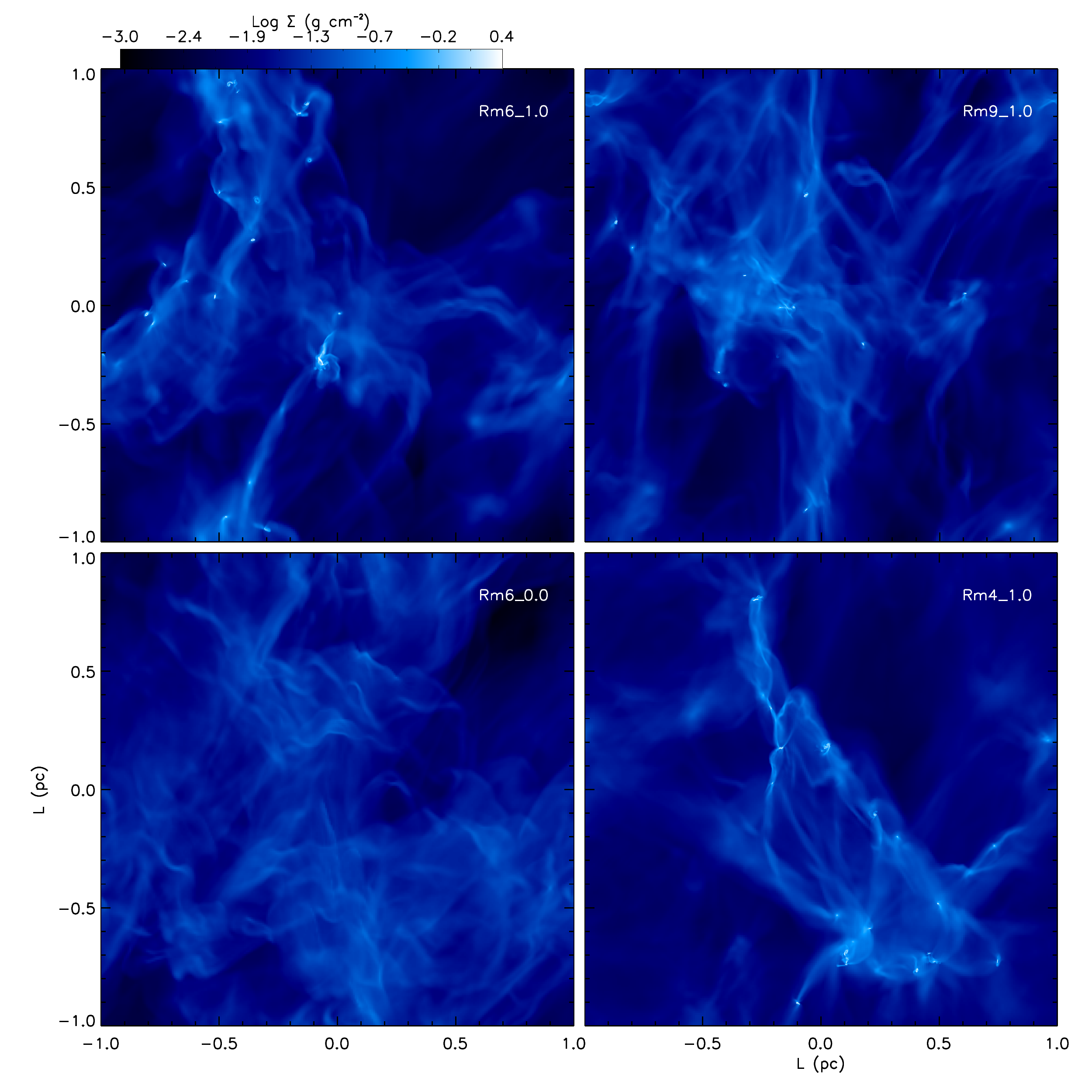}
\caption{Total gas column density for {\sc orion} snapshots Rm6\_1.0 (top left), Rm9\_1.0 (top right), Rm6\_0.0 (bottom left), Rm4\_1.0 (bottom right). The snapshot times are 1$t_{\rm ff}$, 1$t_{\rm ff}$, 0$t_{\rm ff}$, and 1$t_{\rm ff}$, respectively.
\label{column} }
\end{figure*}

\section{Methods} \label{methods}

\subsection{Hydrodynamic Simulations}\label{numsims}

In this paper, we analyze snapshots of four different hydrodynamic simulations of turbulent molecular clouds. The simulation parameters 
are summarized in Table \ref{sim}. Three of the simulations (Rm4, Rm6 and Rm9) are performed with the {\sc orion} adaptive mesh refinement
(AMR) code \citep{truelove98, klein99}. 
%These three simulations do not include
%magnetic fields but are evolved for a free-fall time with large-scale driven turbulence, self-gravity, and sink particles \citep{krumholz04}.
 Since these simulations have not been previously published, we describe our method in detail below. 

{\sc orion} employs a conservative second order Godunov scheme to solve the equations of compressible gas dynamics: 
\beqa
\frac{\partial \rho}{ \partial t} + \nabla \cdot (\rho {\bf v}) &=& 0, \\
\frac{\partial \rho {\bf v}}{ \partial t} + \nabla \cdot (\rho {\bf v v}) &=& -\nabla P - \rho \nabla \phi, \\
{{\partial \rho e}\over {\partial t}} + \nabla \cdot [(\rho e + P ){\bf v}] & = & \rho {\bf v} \nabla \phi, 
\eeqa
where $\rho$, $P$, ${\bf v}$ are the gas density, pressure, and velocity, respectively. Here, $e$ is the total energy $e= \frac{1}{2} \rho {\bf v}^2 + \frac{P}{\gamma -1}$, where $\gamma$ is the ratio of specific heats. {\sc orion} solves the Poisson equation for the gravitational potential, $\phi$:
\beq
\nabla^2 \phi = 4 \pi G \left[ \rho + \sum_n m_n \delta({\bf x}-{\bf x}_n)\right],
\eeq
where $m_n$ and ${\bf x_n}$ are the mass and position of the {\it n}th star, respectively. 

We close these equations with an isothermal equations of state:
\beq
P = \rho \frac{ k_B T}{\mu_p m_{\rm H}},
\eeq
where $k_B$ is the Boltzmann constant, $\mu_p = 2.33$ is the mean mass per particle, $m_{\rm H}$ is the hydrogen mass, and $T = 10$ K is the isothermal gas temperature. Authors sometimes adopt a barotropic equation of state (e.g., \citealt{Offner08a}), which sets a characteristic density above which the gas becomes optically thick and ceases to be isothermal.  However, the density at which this occurs, $\rho_c \sim 10^{-14}$ g cm$^{-3}$, as calculated using full radiative transfer \citep{masunaga98}, exceeds the maximum density at our maximum AMR resolution ($\sim 5 \times 10^{-16}$ g cm$^{-3}$). Consequently, the isothermal approximation is appropriate here. Alternatively, we might solve for the radiation field using a flux-limited diffusion (FLD) approach and thus take into account heating from forming stars \citep{Offner09a}. This would be more numerically expensive but more physically accurate in the dense star-forming gas. However, without some prescription for protostellar outflows the stellar heating in the calculation would be an over-estimate \citep{hansen12}, and moreover, an FLD approach would not supply more accurate information about the temperatures of the low-extinction gas as {\sc 3d-pdr} does.

We insert finer AMR grids when the local density violates the Truelove criterion \citep{truelove97}:
\beq
\rho < \rho_J =  J^2\frac{\pi k_B T}{G \mu_p m_{\rm H} \Delta x_l^2},
\eeq
where $\Delta x_l$ is the cell size on level $l$ and we adopt a Jeans number of $J=0.125$. A sink particle is inserted when the gas exceeds the Jeans density for $J=0.25$ on the maximum level \citep{krumholz04}. In this paper, we do not analyze the sink particle distribution and properties; these are the subject of Kirk et al.~(in preparation).

We initialize the simulations with uniform density and then
perturb the gas for three crossing times using a random velocity
field (e.g., \citealt{maclow99}). This field has a flat power spectrum for wavenumbers $k=1..2$, which corresponds to physical scales of $L..L/2$. We re-normalize the perturbations to maintain a constant
cloud velocity dispersion. In the fiducial simulation, Rm6, the Mach number is chosen to satisfy the observed linewidth-size relation \citep{mckee07}. Following the driving initialization, the simulations 
achieve a well-mixed turbulent state and we turn on gravity, allowing collapse to proceed for a global free-fall time.

The {\sc orion} simulations all have a 256$^3$ base grid and four levels of AMR
refinement. As summarized in Table \ref{sim}, these three calculation have a total gas mass of 600 $\msun$, domain size of 2 pc ($\Delta x_4 = 100$ AU), and turbulent 3D Mach numbers of 4.2, 6.6 and 8.9. For comparison, we also analyze Rm6 without gravity, i.e., at $t=0 t_{\rm ff}$, and at half a free-fall time. Figure \ref{column} shows the integrated column density at one free-fall time for these runs.

We include the third simulation, n300, in order to directly compare our PDR methodology to that of \citet{glover10}, henceforth G10. The n300 simulation was performed by S.~Glover with a modified version of {\sc ZEUS-MP}, which tracks the abundances of 32 chemical species. The n300 calculation uses a fixed 256$^3$ grid. Turbulence is generated using random velocity perturbations in a manner similar to that used for the {\sc orion} simulations. It does not include self-gravity but does solve the equations of ideal magneto-hydrodynamics and begins with an initially uniform magnetic field of 6 $\mu$G.
% add a note about whether the cloud is supervirial.

\begin{figure*}
\epsscale{1.0}
\plotone{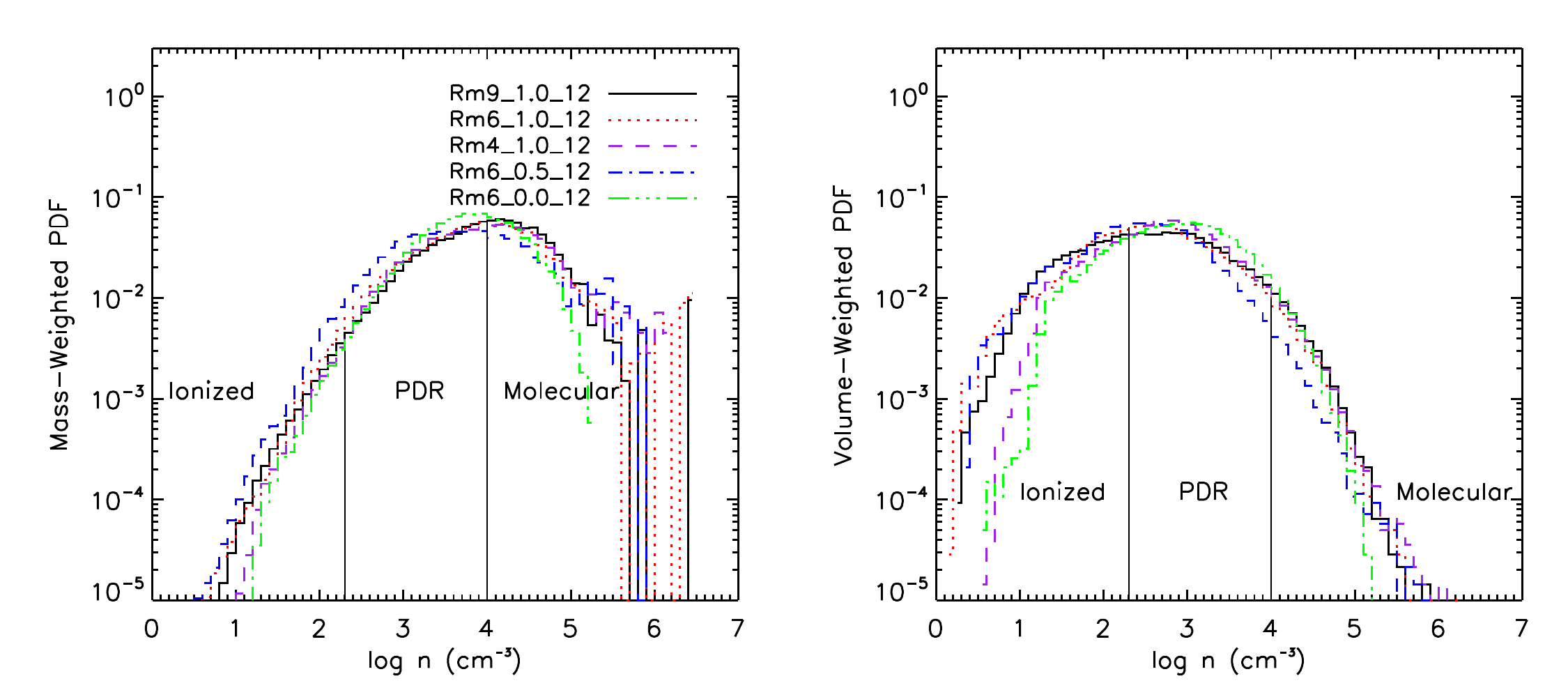}
\caption{Density distributions for runs Rm9\_1.0\_12 (black, solid), Rm6\_1.0\_12 (red, dot), Rm4\_1.0\_12 (purple, dash), Rm6\_0.5\_12 (blue, dot-dash), and Rm6\_0.0\_12 (green, dot-dot-dash). The gas state is characterized as ionized (left), PDR (middle) or molecular (right), where vertical lines indicate the state boundaries.
%\caption{Density distributions for runs Rm9\_1.0\_12 (solid), Rm6\_1.0\_12 (dot), Rm4\_1.0\_12 (dash), Rm6\_0.5\_12 (dot-dash), and Rm6\_0.0\_12 (dot-dot-dash). The gas state is characterized as ionized (purple, left), PDR (red, middle) or molecular (blue, right).
\label{comp_nhist} }
\end{figure*}

Figure \ref{comp_nhist} shows the mass-weighted and volume-weighted density distributions and corresponding chemical regimes for each of the {\sc orion} snapshots. The density distribution functions exhibit a characteristic log-normal shape as expected for supersonic turbulent gas (e.g., \citealt{padoan97,kritsuk07}). As self-gravity becomes important, the density distribution grows a high-density tail \citep{maclow04}. The cells at the peak of the density distribution fall into the PDR regime for the simulation parameters we adopt.
The vertical lines in the histogram indicate the division between ionized, PDR and molecular gas.

\subsection{{\sc 3d-pdr}}\label{3DPDR}
%    -Initial conditions (fiducial problem setup, i.e. uv field implementation, chemical assumptions)

{\sc 3d-pdr} \citep{bisbas12} is a three-dimensional time-dependent astrochemistry code for treating photodissociation regions (PDRs) of arbitrary density distribution. The code is able to solve self-consistently the chemistry and the thermal balance within any three-dimensional cloud. It uses an escape probability approximation \citep[or Large Velocity Gradient --][]{sobolev60, castor70, dejong75} to compute the cooling functions. To do this, {\sc 3d-pdr} uses a ray tracing scheme in which the directions of the rays are controlled by the {\sc healpix} algorithm \citep{gorski05}. This ray tracing scheme creates a discrete set of evaluation points by projecting the elements of the cloud along each ray. It can thus evaluate the column densities, the attenuation of the far ultraviolet radiation into the PDR, and the propagation of the FIR/submm line emission out of the PDR.

As a further development of the fully bench-marked one-dimensional {\sc ucl\_pdr} code \citep{bell06}, {\sc 3d-pdr} adopts the same chemical model features. For the simulations presented in this paper, we use a chemical network which is a subset of the UMIST data base of reaction rates \citep{woodall07}. This ``reduced'' network consists of 320 reactions and 33 species (including electrons). However, \textsc{3d-pdr} also includes heating due to photoionization and photodissociation reactions in addition to the standard gas-phase chemistry. Self-shielding of H$_2$ and CO against photodissociation is accounted for. Comprehensive treatment of various gas heating mechanisms (i.e., photoelectric heating from dust grains and PAHs, collisional de-excitation of vibrationally excited H$_2$ following FUV pumping, photoionization of neutral carbon, cosmic ray heating) and emission from major cooling lines ([CII], [CI], [OI], CO) are calculated at each element.  {\sc 3d-pdr} also includes turbulent heating, which is proportional to $v_{\rm TURB}^3/L$, where $v_{\rm TURB}$ is the turbulent velocity and $L$ is the integral scale. Here, we adopt constant values of $L=5$ pc and $v_{\rm TURB}=1\,{\rm km}\,{\rm s}^{-1}$. In practice, $L$ should be set to the simulation domain size and $v_{\rm TURB}$ to the 1D turbulent Mach number times the mean sound speed, however we find that the turbulent heating is small compared to photoelectric, cosmic-ray and chemical heating, which are the other main sources of heating. (See the Appendix for a discussion of the relative heating rates.)
The thermal balance is solved self-consistently with the chemistry to determine the gas temperature. 
Unless otherwise specified, we adopt total Carbon and Oxygen abundances of $x_{\rm C}=10^{-4}$ and $x_{\rm O}= 3.16 \times 10^{-4}$.
Further details can be found in B12.

For the purposes of this paper we consider as PDR any H-nucleus density within the region $200\le n_{\rm H}\le10^4\,{\rm cm}^{-3}$. Below $n_{\rm H}=200\,{\rm cm}^{-3}$ we consider it ionized, whereas above $n_{\rm H}=10^4\,{\rm cm}^{-3}$ we consider it fully molecular, with constant gas temperature and abundances that are independent of the external radiation field. The lower density limit is somewhat arbitrary since the H to H$_2$ transition can occur down to lower densities depending on the temperature. We impose this cutoff on the PDR calculations since we assume that gas at lower densities represents the HII component of the medium, which can only be reliably modeled using a photoionization code (e.g., MOCASSIN \citet{ercolano03, ercolano05,ercolano08}.

In this paper, once the gas is fully molecular we do not solve for its properties with {\sc 3d-pdr}.  Instead, we adopt the limiting values of the temperature and abundances for a uniform density of  $n_{\rm H}=10^5\,{\rm cm}^{-3}$, which correspond to 10 K  and $n_{\rm CO}/n_{\rm H}=10^{-4}$, wherein no atomic Carbon remains.
%we do not account for the density distribution of the molecular region but instead we treat it as uniform density with $n_{\rm H}=10^5\,{\rm cm}^{-3}$  and 10 K temperature. 
This is a reasonable approximation for these densities since this gas, by definition, is well shielded from the external radiation and is almost entirely molecular. 

The cosmic ray ionization rate per H$_2$ molecule is taken to be $\zeta=5\times10^{-17}\,{\rm s}^{-1}$. 
%The turbulent velocity is fixed to be $v_{\rm TURB}=1\,{\rm km}\,{\rm s}^{-1}$. 
The dust temperature is constant and set to $T_{\rm dust}=20\, {\rm K}$. We use ${\cal N}_{\ell}=12$ rays of {\sc healpix} refinement (level $\ell=0$) and we use $\theta_{\rm crit}=0.5(\simeq\pi/6)\,{\rm rad}$ for the search angle criterion. We neglect the contribution of the diffusive component of the FUV field by invoking the \emph{on-the-spot} approximation \citep{osterbrock74}.  We consider we have obtained thermal balance either when the heating and cooling rates differ by $\sigma_{\rm err}\le0.5\%$, or when the difference in temperature between two consecutive iterations is $T_{\rm diff}\le0.01\,{\rm K}$. 
Finally, we typically evolve the {\sc 3d-pdr} simulation to final times from $5.7-100$ Myr at which point the chemistry is in equilibrium (e.g., \citealt{bayet09}). Table \ref{simpdr} summarizes all the runs we perform with {\sc 3d-pdr}.  %$t_{\rm END}=5.6\,{\rm Myr}$.

Although Rm4, Rm6 and Rm9 each have 4 levels of grid refinement with a minimum cell size of 100 AU, we consider only the $256^3$ base-grid data when post-processing. The refined cloud regions, by construction, contain high-density gas that is $\gtrsim 10^4$ cm$^{-3}$. At these densities,  {\sc 3d-pdr} considers the gas to be fully molecular and adopts a constant gas temperature and abundances.

\begin{deluxetable}{lccccc}
\tablecolumns{6}
\tablecaption{Simulation Properties \label{sim}}
\tablehead{ \colhead{Snapshot ID\tablenotemark{a}} & 
  \colhead{$L$(pc)} & 
  \colhead{$M$($\msun$)} &
  \colhead{$\mathcal{M}$} &
  \colhead{$k_{\rm min}..k_{\rm max}$\tablenotemark{b}} & $t/t_{\rm ff}$} \\
\startdata
Rm6\_0.0 &2 & 600 & 6.6 & 1..2 & 0.0 \\ %74
Rm6\_0.5 &2 & 600 & 6.6 & 1..2 & 0.5 \\ %74
Rm6\_1.0 &2 & 600 & 6.6 & 1..2 & 1.0 \\ %74
Rm9\_1.0 &2 & 600 & 8.9 & 1..2 & 1.0 \\ %13
Rm4\_1.0 &2 & 600 & 4.2 & 1..2 & 1.0 \\ %56
%Rk34 &2 & 600 & 6.6 & 3..4 \\ %14
n300 &20 & 82800  & 12.5 & 1..2 & 0.0 \\%14
\tablenotetext{a}{Simulation output ID, box length, total initial gas mass,
  Mach number, and fraction of a global free-fall time with gravity, respectively.}
\tablenotetext{b}{The wavenumber range of the random velocity perturbations. 
% This is now in the text
%A range of 1..2 should be interpreted as turbulence injected on scales comparable to the size of the cloud to half the size of the cloud.
}
\end{deluxetable}

\begin{deluxetable*}{lcccccc}
\tablecolumns{7}
\tablecaption{{\sc 3d-pdr} Run Parameters \label{simpdr}}
\tablehead{ \colhead{Run ID} & \colhead{Grid ($f\times256^3$)\tablenotemark{a}} & 
  \colhead{FUV (G$_0$)\tablenotemark{b}} & \colhead{Final Time (Myr)} & \colhead{Field Geometry\tablenotemark{c}} & \colhead{${\cal N}_{\ell}$\tablenotemark{d}} & \colhead{$n_l..n_u$ (cm$^{-3}$)\tablenotemark{e}}}  \\
\startdata
Rm6\_0.0\_12 & 1/12 & 1 & 100 & plane-parallel & 12 & $200..10^4$ \\
Rm6\_0.5\_12 & 1/12 & 1 & 100 & plane-parallel & 12  & $200..10^4$ \\
Rm6\_1.0\_12 & 1/12 & 1 & 100 & plane-parallel & 12  & $200..10^4$ \\
Rm6\_1.0\_12i & 1/12 & 1 & 100 & isotropic & 12  & $200..10^4$ \\
Rm6\_1.0\_12ui & 1/12 & $\frac{1}{2}-\frac{1}{2}$ & 100 & plane-parallel + isotropic  & 12  & $200..10^4$ \\
Rm6\_1.0\_25 & 1/25 & 1 & 100 & plane-parallel & 12  & $200..10^4$ \\
Rm6\_1.0\_50 & 1/50 & 1 & 100 & plane-parallel  & 12  & $200..10^4$\\
Rm6\_1.0\_12b\tablenotemark{f} & 1/12 & 1 & 5.7 & plane-parallel & 12 & $200..10^4$ \\
Rm6\_1.0\_12\_48 & 1/12 & 1 & 100 & plane-parallel & 48  & $50..10^4$ \\
Rm6\_1.0\_12\_NC & 1/12 & 1 & 100 & plane-parallel & 48  & $0..10^4$ \\
Rm9\_1.0\_12 & 1/12 & 1 & 100 & plane-parallel & 12  & $200..10^4$  \\
Rm4\_1.0\_12 & 1/12 & 1 & 100 & plane-parallel & 12  & $200..10^4$ \\
n300\_12\tablenotemark{f} & 1/12 & 1 & 5.7 & plane-parallel & 12 & $200..10^4$\\
\enddata
\tablenotetext{a}{Input sampling of the simulation data used by {\sc 3d-pdr}.}
\tablenotetext{b}{Magnitude of the UV field in Draines at the box edge.}
\tablenotetext{c}{The direction of the field at the boundaries. The field is either a uniform field that is plane-parallel to the box faces, isotropic, or a combination of the two.}
\tablenotetext{d}{Number of rays.}
\tablenotetext{e}{Range of {\sc 3d-pdr} densities assumed in the calculation.}
\tablenotetext{f}{Run uses the same C and O abundances as G10 ($x_C = 1.41\times10^{-4}$ and $x_O=3.16 \times 10^{-4}$).}
\end{deluxetable*}

%? separate section or in 3d-pdr above?
\subsection{``One-Way'' Hydrodynamic-Chemical Coupling}

Our method can be considered a ``one-way'' code coupling, because {\sc 3d-pdr} uses the density output of the hydrodynamic calculations to compute the chemical distribution. A benefit of this approach is that it is computationally efficient, and large networks of reactions may be considered that would otherwise be too time consuming to compute in combination with the hydrodynamics. In addition, the affects of different radiative conditions and metallicity may be studied using the same hydrodynamical simulation. The deficit to this approach is that the corresponding temperatures computed by {\sc 3d-pdr} do not affect the subsequent hydrodynamic evolution. In a one-way coupling, consistency between the hydrodynamic quantities and chemistry is only achieved if the a priori simulated values are chosen to reflect the anticipated post-processed values.  Because {\sc 3d-pdr} computes a wide distribution of temperatures, it is not possible to achieve consistency by adopting a single, constant temperature. For example, for Rm6\_1.0\_12 {\sc 3d-pdr} determines a mass-weighted temperature of $\sim$22 K, which is a factor of two above the fiducial 10 K simulation temperature.  However, because we adopt 10 K for the simulation, by construction the densest regions, i.e., the star-forming gas ($ n \gtrsim$ a few $10^3$), their dynamics will be in fairly good agreement with the computed {\sc 3d-pdr} temperatures. It is also worth noting that for a simulation with a 1D rms velocity of 0.7 km s$^{-1}$, gas temperatures would need to reach $\sim$ 140 K in order to obtain dynamic parity with the turbulent gas pressure (assuming a stellar external radiation field). Since the {\sc 3d-pdr} computed gas temperatures are generally much less than 140 K, the hydrodynamics would remain governed by turbulence and so only a small difference would be expected if the {\sc 3d-pdr} temperatures were fed back into the simulation.

In the simulation we also adopt a fixed value for the mean mass per particle, $\mu_p$, which implicitly assumes that the gas is entirely molecular. We will show later that the hydrogen is almost all in molecular form throughout the domain with the exception of a few cells at the domain edge. Since molecular hydrogen dominates the mass budget of the gas by several orders of magnitude this particle mass approximation is a good one for the simulations used in this study.

A second discrepancy between the dynamics and the chemistry occurs because {\sc 3d-pdr} assumes that the radiation field impinges on the gas at the box boundaries, while the hydrodynamics assume periodic boundary conditions, i.e., there is no edge. This incongruity is also part of the G10 approach, which adopts periodicity for the gas but not the radiation field. For any boundary convention, high-density gas will have high-extinction nearly independently of location with respect to the boundary. Since turbulent clouds are naturally porous and the dense gas has a low-volume filling fraction, we can expect that radiation would penetrate many lower density regions for some sight-line to the ``edge.'' Practically, the effect of the incident radiation field is to define a new effective boundary for the molecular gas, which reflects the filamentary and inhomogenous shape of the gas. Authors that seek to model an entire cloud rather than a periodic piece must instead wrestle with the arguably equally difficult problem of how the cloud connects to the larger-scale ISM, which is related to the issue of molecular cloud formation (e.g., \citealt{banerjee09, vanloo13}).

\section{Method Validation}\label{validation}

%Note--these files do not contain rho so can't do mass weighted abundances here.
\begin{deluxetable*}{lcccccc}
\tablecolumns{7}
\tablecaption{Mean fractional values at various resolutions \label{conv}}
\tablehead{ \colhead{Snapshot ID\tablenotemark{a}} & 
  \colhead{H} & 
  \colhead{H$_2$} &
  \colhead{C+} &
  \colhead{C} & CO & T (K)} \\
\startdata
Rm6\_1.0\_12 & 0.082 (0.16)& 0.36 (0.20)  & 6.7d-5 (4.0d-5) & 6.9d-6 (1.0d-5)  & 5.24d-6 (1.5d-5) & 45.7 (32.2)  \\ 
Rm6\_1.0\_25 & 0.099 (0.19)& 0.35 (0.20) & 6.7d-5 (4.0d-5) & 6.8d-6 (1.0d-5) & 5.0d-6 (1.5d-5) & 45.5 (31.8)\\
Rm6\_1.0\_50 & 0.10 (0.21)& 0.34 (0.20)  & 6.6d-5 (4.1d-5) & 6.6d-6 (9.9d-6) & 5.1d-6 (1.5d-5) & 44.1 (31.9)
\tablenotetext{a}{Simulation output ID and mass-weighted mean abundances. The standard deviation for each is given in parentheses.}
\end{deluxetable*}

\subsection{Grid Sampling}   

We first verify that our results are converged and independent of the {\sc 3d-pdr} grid resolution by comparing the calculated abundances for the same simulation input (Rm6\_1.0) sampled with three different resolutions. 
%SSRO
These are the runs Rm6\_1.0\_12, Rm6\_1.0\_25 and Rm6\_1.0\_50 listed in Table \ref{simpdr}. 
This is a useful exercise because {\sc 3d-pdr} post-processing requires non-negligible time even when run in parallel. Throughout this paper, we analyze a coarser resolution than is actually achieved by the hydrodynamic simulations. 

Table \ref{conv} gives the mean abundance and standard deviation over all grid points for each of the three sampling resolutions. We find that differences in the mean abundances are generally only a few percent and are, without exception, much smaller than the standard deviation of the distributions. The mean gas temperature is also fairly insensitive to increasing resolution.

\begin{figure}
\epsscale{1.2}
\plotone{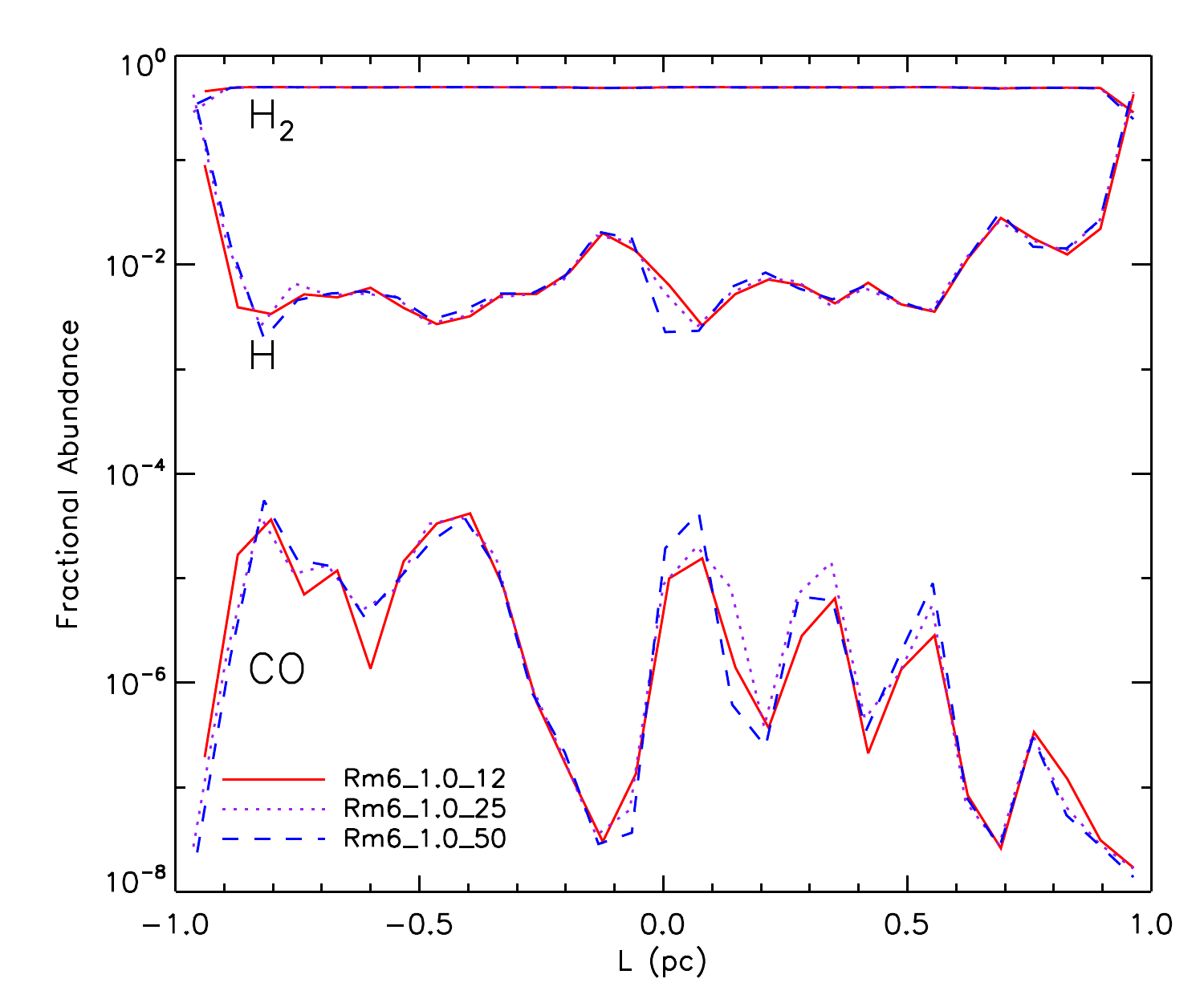}
\caption{Mean fractional H, H$_2$ and CO abundances for a line of sight through the cloud center at three different resolutions. The resolutions differ by factors of two in the number of grid points. 
\label{1D_res} }
\end{figure}

Figure \ref{1D_res} shows the fractional abundances for a single random sight-line through the cloud. Increasing the sampling resolution of {\sc 3d-pdr} has little effect on the calculated cloud chemistry and the abundances of H, H$_2$ and CO. Different sight-lines exhibit similar good convergence.
The small differences between resolutions imply that the results should also be similar for simulation data with higher base grid resolutions. This comparison suggests that in the future it will be possible to follow the time-dependent chemical evolution coarsely but accurately with {\sc 3d-pdr}. However, for stronger UV fields, the resolution could be more important since the C$^+$/C/CO transition will occur further from the boundary.
%SSRO Tom didn't specify if he wanted to comment here or in the section below.

\subsection{Boundary Convergence}

\begin{figure}
\epsscale{1.25}
\plotone{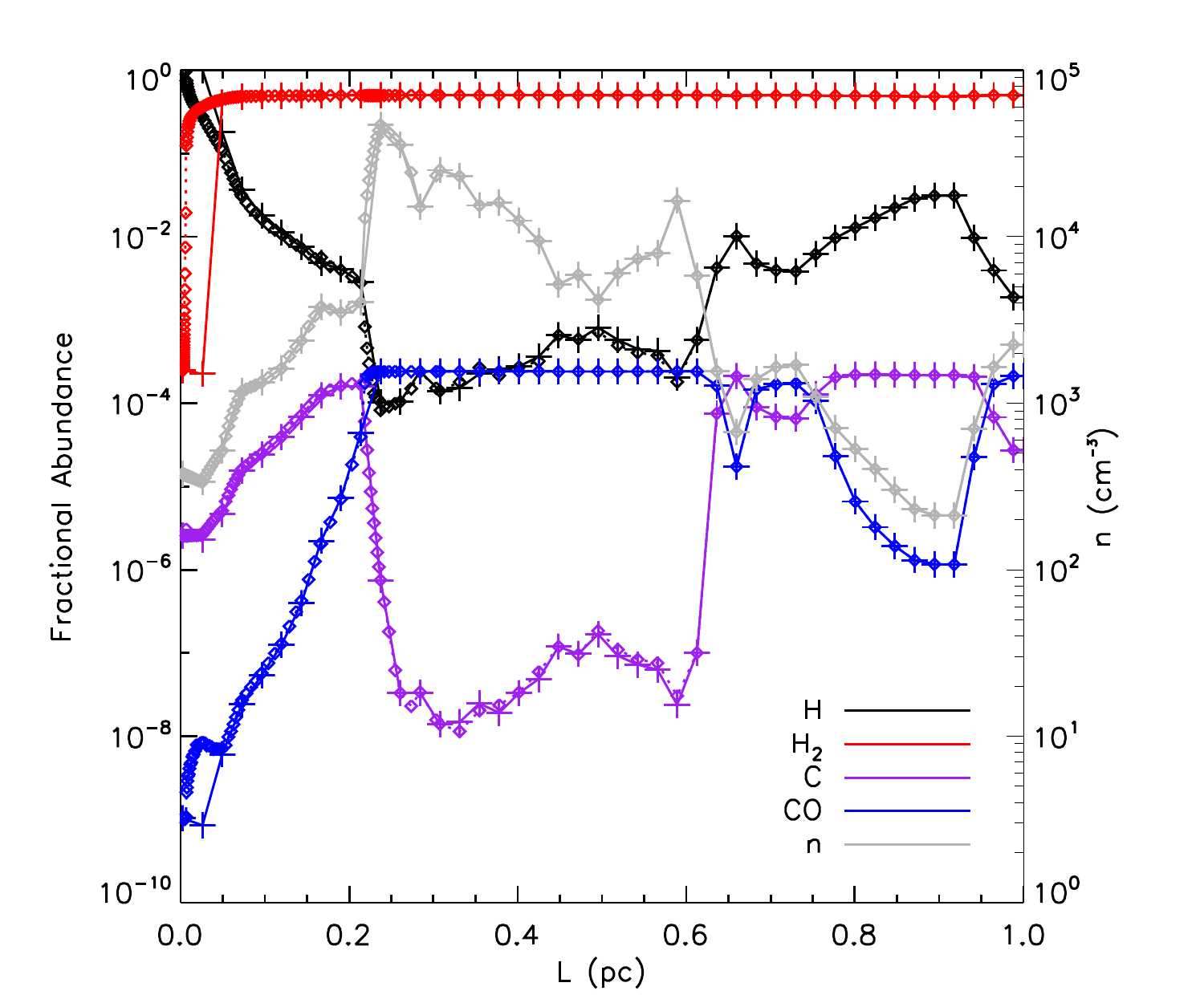}
\caption{Mean fractional H, H$_2$, C, and CO abundances for a line of sight through the cloud center for constant resolution (crosses with a solid line) and for logarithmic spacing at the cloud edge (diamonds with a dotted line). The right axis indicates the gas number density (gray). A uniform 1 Draine radiation field is imposed on the cloud from the left ($L=0$ pc).
\label{edge} }
\end{figure}

Some authors have suggested that the details of the interior cloud chemistry depend on the resolution of the atomic-to-molecular transition. To investigate this issue, we compare  molecular abundances in the cloud interior for two cloud edge resolutions. Figure \ref{edge} shows the same sight-line computed with a fixed linear spacing and with logarithmically spaced points concentrated at the boundary. 
All grid points are assumed to be part of the PDR and are treated with the PDR code.
We find that the abundances in the cloud interior are virtually identical despite the very different boundary resolutions. In fact, the values computed with coarse resolution vary somewhat {\it only} within one or two coarse cells directly adjacent to the boundary. This demonstrates that the chemistry in the cloud interior is not sensitive to the edge resolution for the densities and FUV field strengths considered here and provides further evidence that our lower resolution {\sc 3d-pdr} calculations are chemically converged for the bulk of the cloud.

\subsection{Ray Convergence}\label{rayconv}

\begin{figure}
\epsscale{1.25}
\plotone{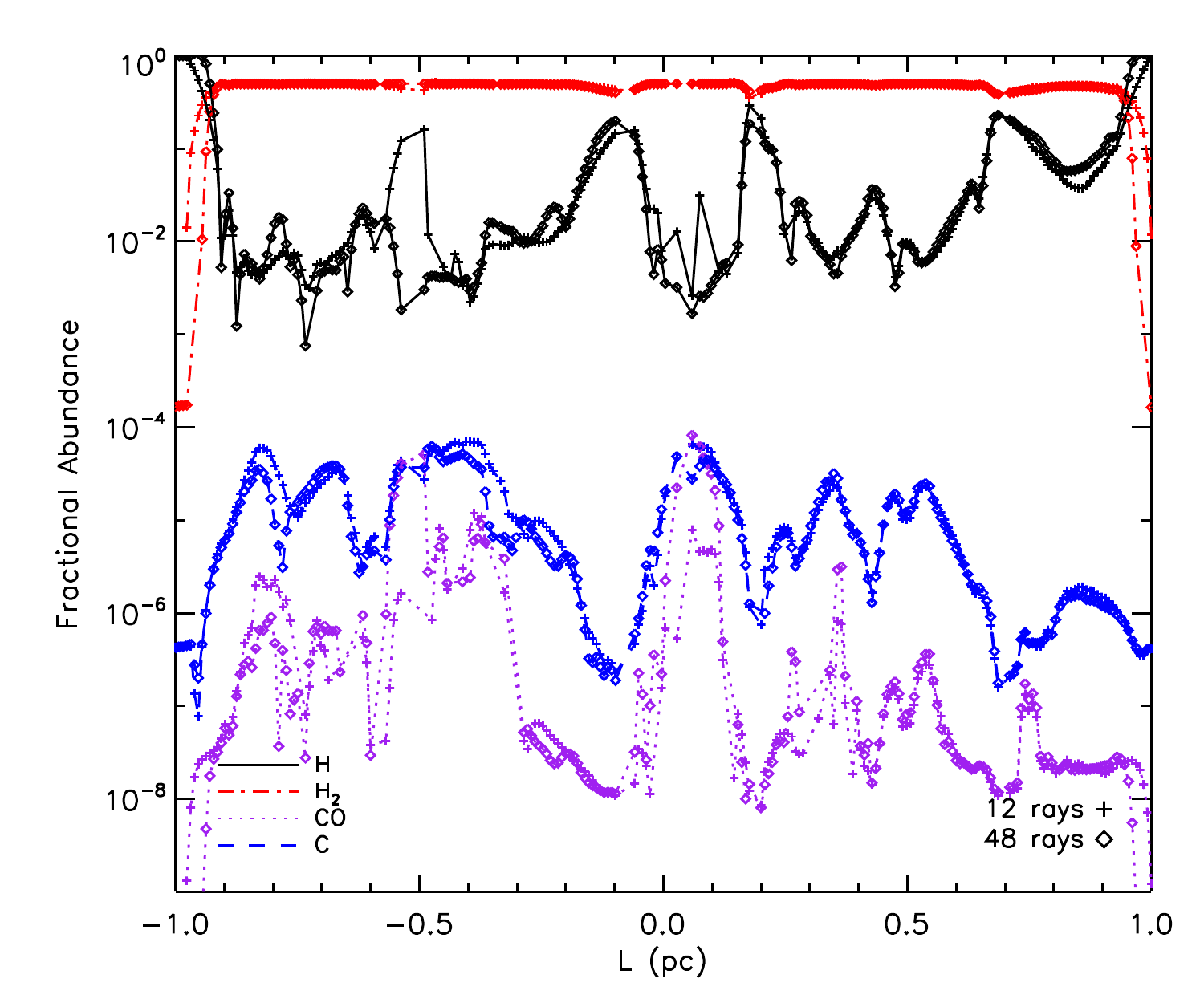}
\caption{Mean fractional H (solid), H$_2$ (dot-dashed), C (dashed), and CO (dottted) abundances for a line of sight through the cloud center for Rm6\_1.0\_12\_NC (crosses) with the fiducial ray resolution (${\cal N}_{\ell}=12$) and for Rm6\_1.0\_12\_48 (diamonds), which has ${\cal N}_{\ell}=48$. Here we plot only the abundances for $50 \le n \le 10^4$ cm$^{-3}$, which are the {\sc 3d-pdr} density limits of Rm6\_1.0\_12\_48.
\label{ray} }
\end{figure}

In order to assess the sensitivity of our results to the number of rays, ${\cal N}_{\ell}$, we compare {\sc 3d-pdr} calculations with 12 ($l=0$) and 48 ($l=1$) rays. In principle, higher ray resolution will be more accurate for asymmetric and fractal geometries. Figure \ref{ray} shows the fractional abundances for a line of sight through the cloud center. Generally, we find good agreement for the two resolutions.  The H$_2$ and C abundances are almost identical, while some differences of up to an order of magnitude are apparent for some H and CO points. For H$_2$ and CO, the resolution does affect the molecular transition at the boundary, where the abundance is lower at higher ray resolution. We can understand this by considering the simpler 6-ray case for a cell on the domain boundary. Assuming that no radiation impinges on the cell from the opposite cloud edge, this cell should see 2$\pi$ sr of the UV field and be completely unshielded. However, for 6 perpendicular rays, only the ray perpendicular to the boundary will see the UV field, which results in an angular attenuation of $4 \pi/6 = 2\pi/3$ sr. Depending on the field strength, this may be sufficient to shield the boundary cell from the UV field. As more rays are added the angular dependence of the field at the boundary becomes better resolved, reducing the amount of extinction. In Figure \ref{ray}, we see this issue only affects a few cells adjacent to the domain edge and does not appear to directly impact the subsequent internal cloud chemistry. 

% Need to change this from 24->48 and change line strengths

\section{Results}\label{results}

\subsection{Code Comparison: Post-processing vs. In situ Calculation}

In this section we compare our results using {\sc 3d-pdr} to the coupled chemical and dynamical method described in G10.
There are a few key differences between the two approaches. {\sc 3d-pdr} follows 320 reactions of 33 species (including electrons) while G10 follows 218 reactions of 32 species. We note that these 218 reactions are not an exact subset of the 320 followed by {\sc 3d-pdr} since they include more reactions with negative ions. G10 adopts the older reaction rates of UMIST99 \citep{leteuff00} instead of UMIST07 \citep{woodall07}. G10 employs a ``six-ray'' approach \citep{nelson97, nelson99, glover07b} to calculate the local attenuated radiation field whereas {\sc 3d-pdr} uses ${\cal N}_{\ell}=12 \times 4^{\ell}$ rays (in this paper we use 12 rays, i.e. $\ell$=0).
%, which provides slightly better accuracy of the local extinction. 
Both methods include heating due to the photoelectric effect, H$_2$ photodissociation, UV pumping of H$_2$, H$_2$ formation on dust grains, and cosmic ray ionization. However, {\sc 3d-pdr} also includes photo-ionization of neutral Carbon and turbulent heating. 

%SSRO
Both methods neglect the impact of the gas velocity distribution on the chemistry. In practice, the details of the velocity field affect the H$_2$ shielding, since the H$_2$ photodissociation rate from any given Lyman-Werner line is related to the escape probability for that line (see \citet{glover07a} and discussion therein). G10 and previous papers instead adopt a six-ray approximation to estimate the shielding, which includes no velocity information. {\sc 3d-pdr} relates the line optical depth to an effective linewidth, which is proportional to the root mean square of the thermal sound speed and turbulent gas velocity. 
%employs a simplified large velocity gradient method, which assumes that the line profile is gaussian with width proportional to the root mean squares of the thermal sound speed and turbulent gas velocity. This means that 
%

In modeling cooling, both methods include emission by C, C$^+$ and O fine structure lines, gas-grain collisional cooling, cooling by rotational lines of CO, and H$_2$ collisional dissociation, but {\sc 3d-pdr} neglects H$_2$ and H$_2$O ro-vibrational and OH rotational lines, which are included in the one-dimensional code {\sc ucl\_pdr} \citep{bayet10}, as well as H collisional ionization and Compton cooling. However, these lines do not produce significant cooling under the conditions considered here, and so neglecting them is a good approximation.
%See 2.6 in Bisbas--how is turbulence included?

To perform a precise comparison of the two methods we apply {\sc 3d-pdr} directly to the density field of n300. We adopt identical C and O abundances ($x_{\rm C}=1.41\times 10^{-4}$ and $x_{\rm O}=3.16\times 10^{-4}$ in all forms relative to hydrogen), evolve to the same final time of 5.7 Myr, and apply the same stellar FUV field at the boundary. 
%SSRO
%Since G10 does not distinguish between ionized and PDR gas, we do not apply the 200 cm$^{-3}$ minimum cutoff for PDR that we adopt above.
%

\begin{figure*}
\epsscale{1.0}
\plotone{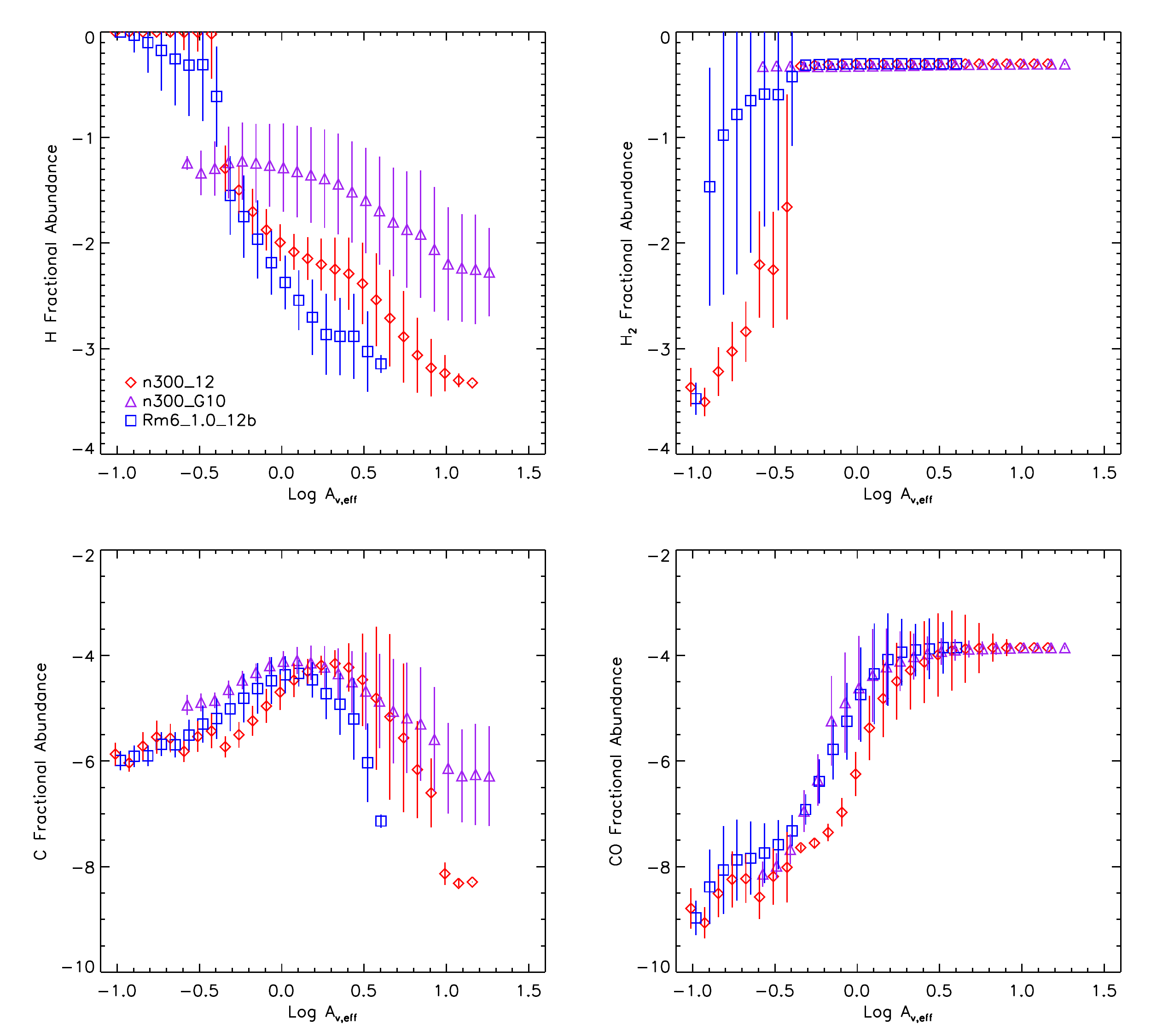}
\caption{Mean fractional H, H$_2$, C, and CO abundances as a function of log extinction, where the error bars indicate the standard deviation in each bin. The triangles show the results of the n300 simulation as performed by S.~Glover. The diamonds indicate the results for {\sc 3d-pdr} assuming the same n300 density distribution evolved to the same time (5.7 Myr) using the same $x_C$ and $x_O$ abundances. The squares indicate the results for Rm6\_1.0\_12b, which has a different underlying density distribution. In all cases, only gas with $200$ cm$^{-3} \ge n \le 10^{4}$ cm$^{-3}$ is included in the averages.
\label{comp_dir} }
\end{figure*}

\begin{figure*}
\epsscale{1.0}
\plotone{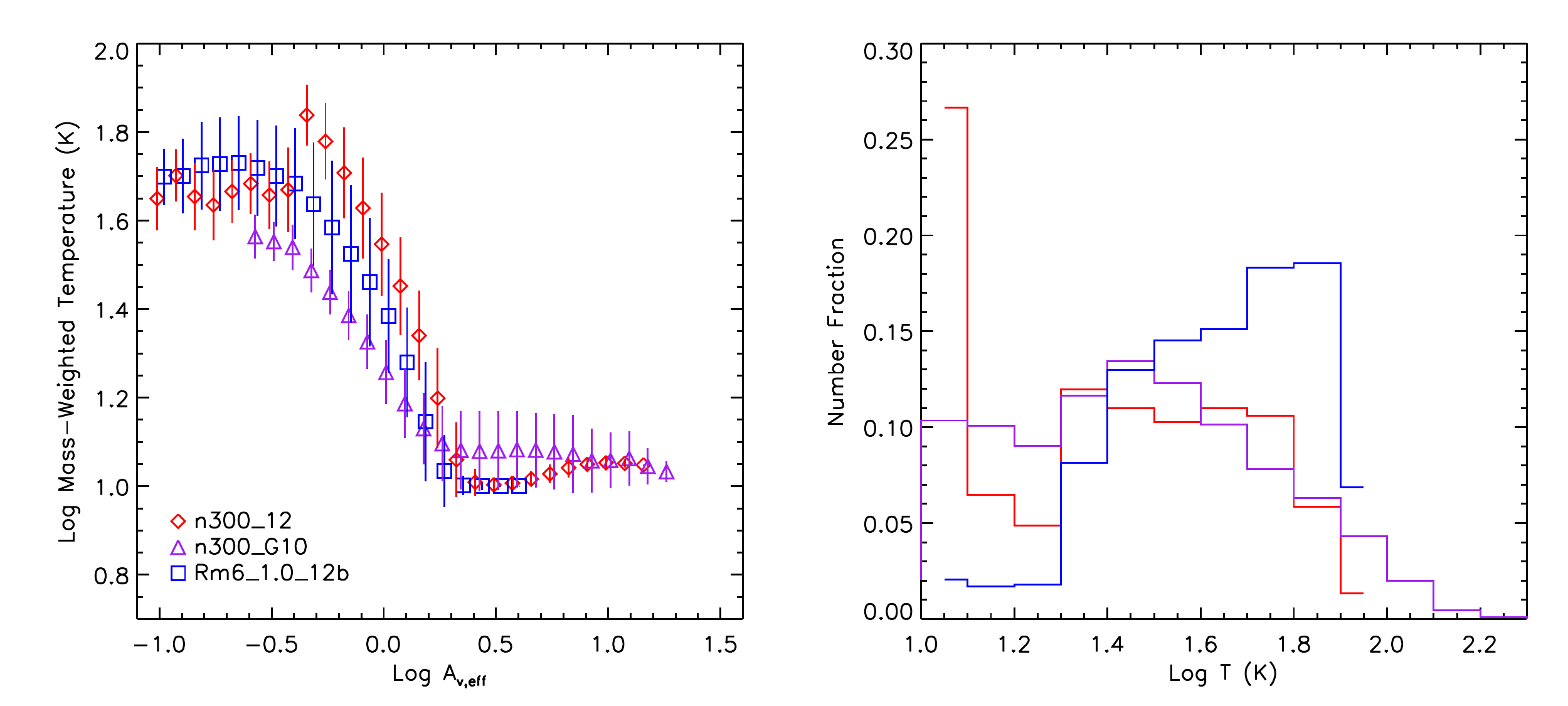}
\caption{Mass-weighted temperature distribution versus extinction (left) and temperature distributions (right) for the same runs shown in Figure \ref{comp_dir}.
\label{comp_dir_temp} }
\end{figure*}

Figure \ref{comp_dir} shows the H, H$_2$, C, and CO abundances binned as a function of effective extinction for the two methods.  We define the effective extinction following G10:
\beq
A_{\rm v, eff} = -\frac{1.0}{2.5} {\rm ln} \left(\frac{1}{{\cal N}_{\ell}} \sum_{i=1}^{{\cal N}_{\ell}} e^{-2.5A_{\rm v}[i]}\right), 
\eeq
where $A_{\rm v, eff}$ is a weighting over the extinctions of all ${\cal N}_{\ell}=12$ rays. For comparison, we include the abundances for a different simulated density field in order to assess the sensitivity to the underlying density distribution.  In Figures \ref{comp_dir} and \ref{comp_dir_temp} we consider only densities $200$ cm$^{-3} \le n \le 10^4$ cm$^{-3}$.\footnote{We find that Figures \ref{comp_dir} and \ref{comp_dir_temp} appear very similar assuming a lower cutoff of $n \ge 50$ cm$^{-3}$ (e.g., Rm6\_1.0\_12\_48 and Rm6\_1.0\_12\_NC). The main result of including lower densities in the PDR calculation is that the low-Av gas becomes increasingly (and inaccurately) warm.}

We find that the G10 and {\sc 3d-pdr} results differ the most at low extinction ($A_{\rm v, eff} <0.3$). These cells are near the simulation boundaries, where the impinging radiation field dissociates the H$_2$. Surprisingly, this transition is largely absent in the G10 method, which also appears to over-estimate the fraction of atomic hydrogen throughout the PDR. At higher extinction, the G10 and {\sc 3d-pdr} methods show reasonable confluence between the distributions of C and CO. The similarity between the n300 and Rm6\_1.0\_12 distributions illustrates that the PDR is not overly sensitive to the underlying density distribution. 

We note that the abundance distributions of Rm6\_0.0\_12, which we evolve to 100 Myr, and those of Rm6\_0.0\_12b, which we conclude at 5.7 Myr, are very similar. This suggests that these species achieve chemical equilibrium by 5.7 Myr (see also \citealt{bayet09}). The formation time of molecular hydrogen for gas with $n=10^{3}$ cm$^{-3}$ is $t_{\rm form} \simeq 10^9\,n^{-1}  {\rm ~yr} \simeq $ 1 Myr \citep{hollenbach71}. CO forms rapidly provided $A_{\rm v} \gtrsim 0.7$ \citep{bergin04}
%SSRO What is the formation time of CO for n=300 gas?

Figure \ref{comp_dir_temp} illustrates the gas temperature distributions in the three cases. G10 reaches a slightly lower temperature at low Av, but otherwise the calculations are within a standard deviation. The temperature histograms, which show the relative number of cells at different temperatures, are somewhat different. 
%The figure shows the two {\sc 3d-pdr} cases only for densities below the fully molecular $10^4$ cm$^{-3}$ limit. 
All simulations exhibit a peak in the temperature distribution at $\sim 30-50$ K.

\begin{figure}
\epsscale{1.3}
\plotone{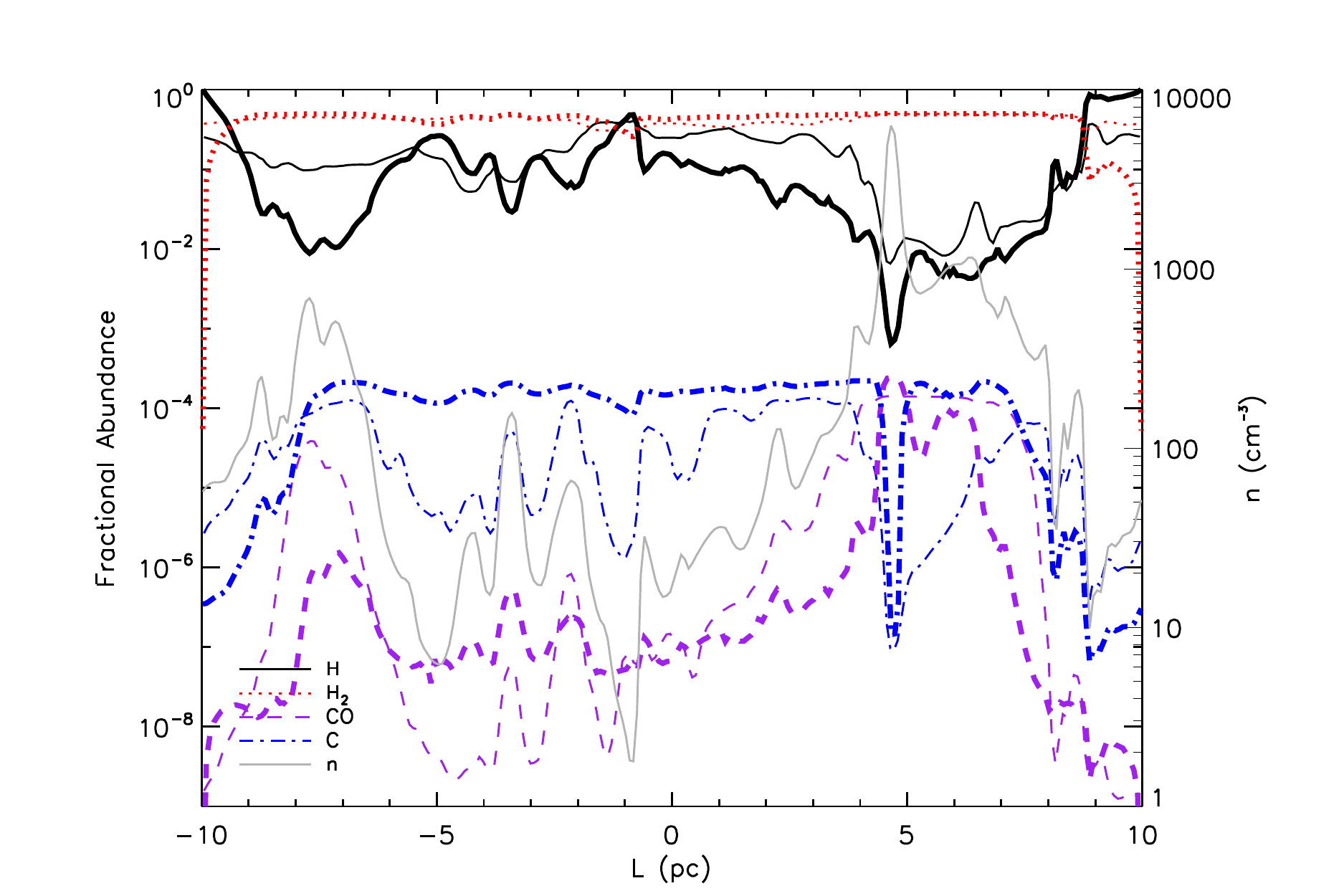}
\caption{Fractional abundances computed for a line-of-sight through the cloud center of simulation n300 for the G10 method (narrow lines) and {\sc 3d-pdr} (thick lines). The gas number density (right axis) is indicated by a gray, solid line.  For comparison, no cutoff is applied to the {\sc 3d-pdr} calculation.
\label{1d_comp} }
\end{figure}

Despite the general congruence of mean properties, Figure \ref{1d_comp} demonstrates that individual cells may have very different fractional abundances. H$_2$ abundance has the best point-by-point agreement since it is nearly constant throughout the domain. The exception occurs at the cloud boundaries, where the G10 method does not appear to model the PDR regime as well as {\sc 3d-pdr}. The discrepancy is likely due to the lower ray resolution of G10, which causes the H$_2$ fraction to be over-estimated (see discussion in section \ref{rayconv}). G10 mention ray resolution as a possible deficit of their 6-ray approach. The H abundance shows fairly good correspondence between the two methods but does differ by an order of magnitude at some points. The higher H fractional abundance at high-Av shown for n300\_G10 may be due to turbulent mixing, which we do not include in our approach. However, since n300 does not include gravity, which would reduce turbulent turnover at high-densities, this H fractional abundance may also be an over-estimate (S. Glover private communication).  The C and CO abundances produced by the two methods are also generally similar with the most difference occurring in the range $-5 {\rm ~ pc} <x <0$ pc, which corresponds to gas densities $n< 10^2$ cm$^{-3}$. 

Since the input densities, grid resolution, and atomic abundances are identical, all discrepancies must be due to differences in methodology and chemical assumptions. Although the magnitude of the abundance variation appears quite large, such differences are consistent with those typically found between PDR codes \citep{rollig07}.

%SSRO Note: If I made it two column it took up a whole page. Leave as is for editors to fix.

\begin{figure}
\epsscale{1.2}
\plotone{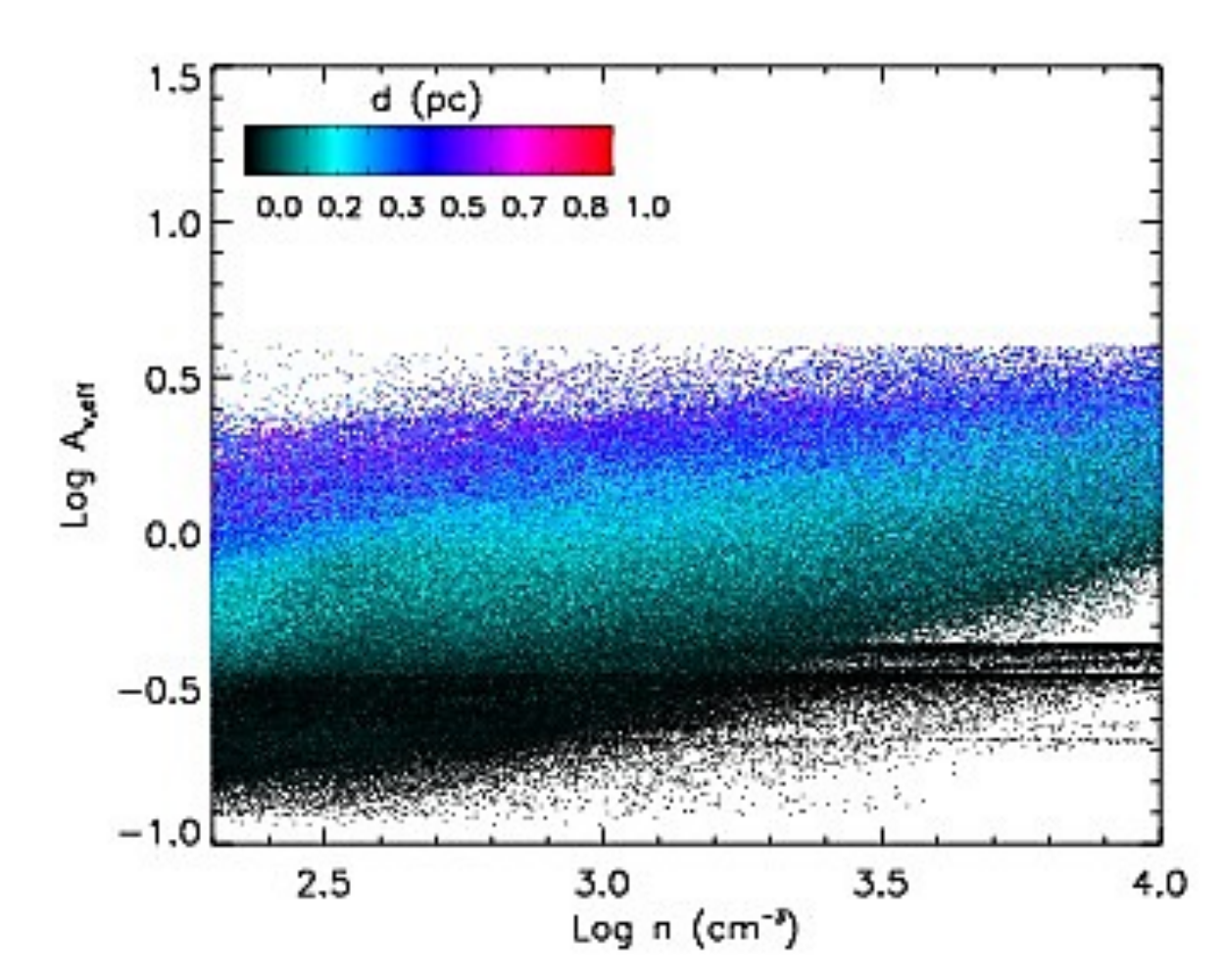}
\caption{Log extinction versus number density for Rm6\_1.0\_12. The colorbar indicates the minimum distance to a cloud boundary in pc.
\label{av_vs_n}}
\end{figure}

\begin{figure*}
\epsscale{0.9}
\plotone{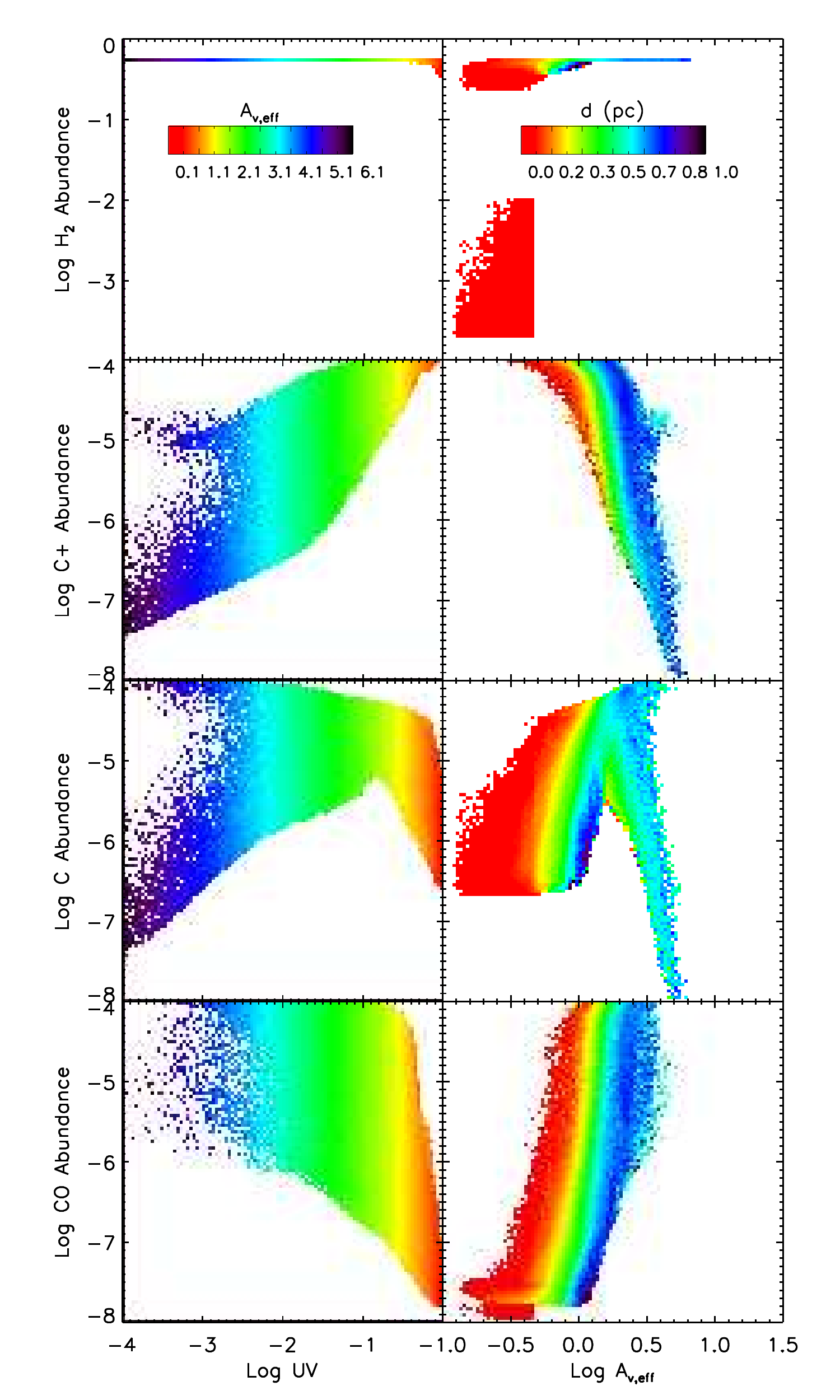}
\caption{Left: Rm6\_1.0\_12 fractional abundances as a function of log UV radiation field where the colorbar indicates the extinction. Right: Rm6\_1.0\_12 fractional abundances as a function of log extinction (right), where the colorbar indicates the minimum distance in pc to the cloud boundary.
\label{abun_vs_av}}
\end{figure*}

\begin{figure*}
\epsscale{0.9}
\plotone{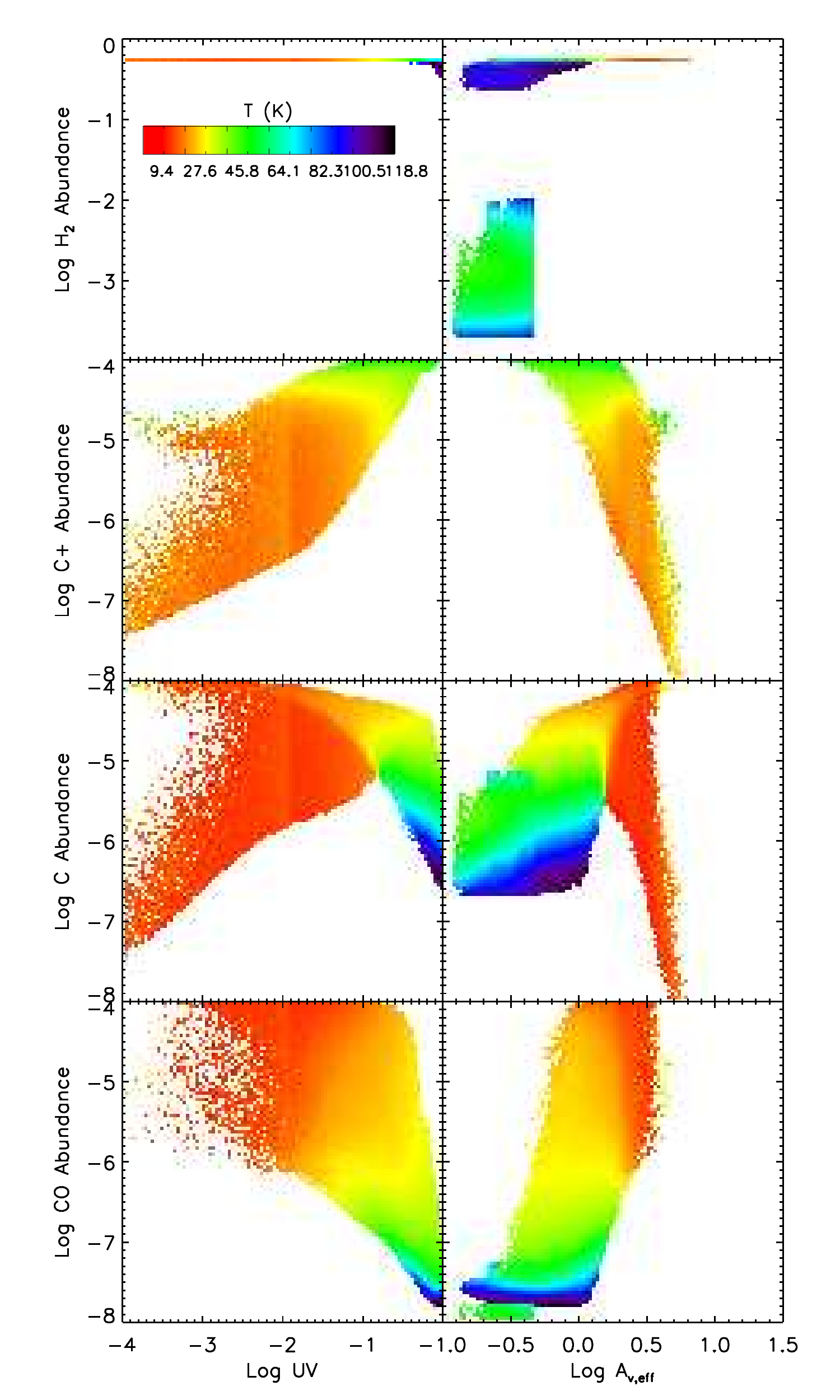}
\caption{Same as Figure \ref{abun_vs_av} but with color indicating the gas temperature.
\label{abun_vs_temp}}
\end{figure*}

\begin{figure*}
\epsscale{0.9}
\plotone{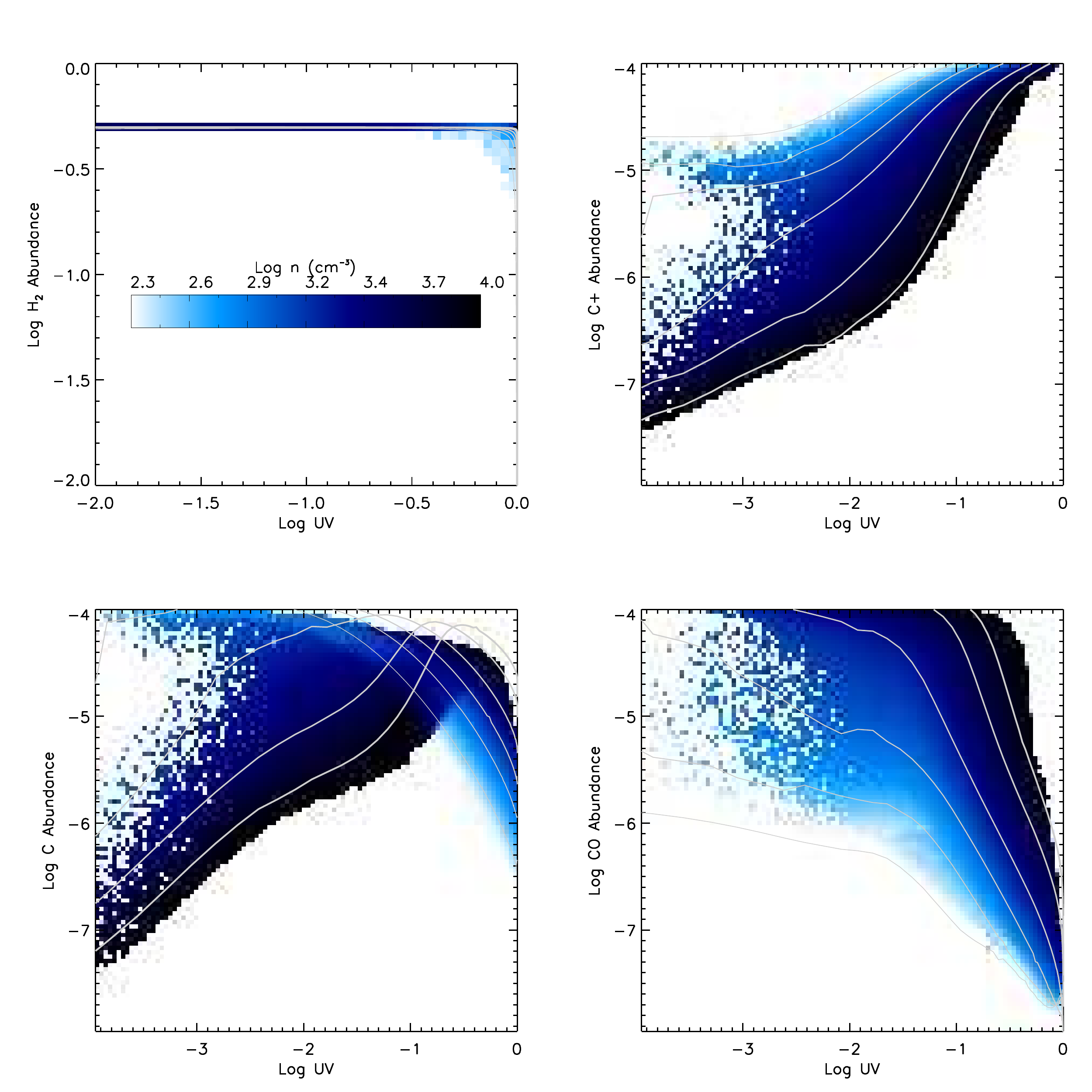}
\caption{Log abundance versus log UV field for run Rm6\_1.0\_12, where the color scale indicates the log of the mean density for a given abundance and local field. The gray solid lines show the values from 1D runs with constant density and incident 1 Draine UV field. The lines increase in density for line thickness going from thin to thick with values of 200, 500,10$^3$, $2\times10^3$, $5\times10^3$, 10$^4$ cm$^{-3}$.  
\label{den_range}}
\end{figure*}

\subsection{Dependence on Dimensionality: 1D versus 3D}

Figure \ref{av_vs_n} shows the extinction distribution as a function of the local gas number density. As shown by G10 (e.g., their Figure 14), gas extinction and density are only very weakly correlated. Overall, extinction is more strongly correlated with position within the cloud than with density.  Figure \ref{av_vs_n} exhibits two populations of points: those in the interior and those spatially within $<1\%$ the boundary, which have distinctly lower extinction and stand out in the regime $3.5< {\rm log} n < 4$. These boundary cells appear to be missing from the results of G10. We discuss them further in section \ref{dep_phys_param}. 

Figure \ref{abun_vs_av} illustrates how the H$_2$, C$^+$, C, and CO fractional abundances depend upon extinction, UV field, and location within the cloud. Both H$_2$ and CO, which are the most sensitive to the extinction, appear to behave differently at the cloud edge and in the interior. Some of the discontinuity in the distribution is likely artificial since better edge resolution, as shown in Figure \ref{1D_res}, would join the two populations more smoothly. 

The left column plots verify that the the local extinction and UV field magnitude are completely correlated (i.e., the extinction indicated by the color-scale varies completely linearly with the magnitude of the UV field). Relative to extinction, proximity to the cloud edge has a weaker influence since the abundances depend upon the column density along each ray, which varies as the density distribution.

Figure \ref{abun_vs_temp} illustrates how abundances correlate with the gas temperature. Temperature varies smoothly with both UV and $A_{\rm v, eff}$. For C, a discrete region of boundary cells becomes apparent, which was previously degenerate with the cells near but not abutting the boundary. These boundary cells show up as a slight offset in temperature for $A_{\rm v, eff}< -0.5$.

%SSRO
The shape of the abundance-UV field distributions depends on the range of underlying densities in the PDR. Figure \ref{den_range} shows the abundance distributions over plotted with lines showing the abundances computed for simple 1D models. The 1D models assume constant density along the line-of-sight and an incident 1 Draine UV field at one end. These curves illustrate that the range in abundance for any given UV field simply depends on the range in local gas density for a given UV field. Since we define the PDR region as those points with densities $n=200$ cm$^{-3}$ to 10$^4$ cm$^{-3}$ the curves with these densities correlate well with the data of the 3D simulations. Thus, while turbulence dictates the distribution of densities and hence the fraction of cells within a given density range, the abundance distribution is set by chemistry and the details of the species response to the local UV field. In summary, although the determination of the extinction at a particular point within the volume is a three-dimensional problem, we find that once the local UV field is computed, the resulting abundances are nearly identical to those derived from a 1D model.

\subsection{Dependence on Physical Parameters}\label{dep_phys_param}

In this section we investigate the sensitivity of the chemistry to the bulk simulation properties. In a self-consistent treatment of molecule formation, the distribution of shock properties (e.g. the post-shock densities and temperatures) could imprint an observable signature in the measured abundances. For example, molecules such as CH$_2$, HCO$^+$ and OH are directly sensitive to turbulent density fluctuations \citet{kumar13} and would likely vary as a function of simulation Mach number. In contrast, the abundance of H$_2$ and CO predominantly depends on the amount of shielding from the UV radiation field \citep{bergin04}. A parcel of gas embedded within a completely smooth ($A_v >0.7$) cloud will be well-shielded from the UV field, whereas a parcel of gas in a highly fractal cloud will have a high probability of having a sight-line with low extinction. Consequently, we can expect that the morphological distribution  of the gas will have some effect on these abundances. 

\begin{figure*}
\epsscale{1.0}
\plotone{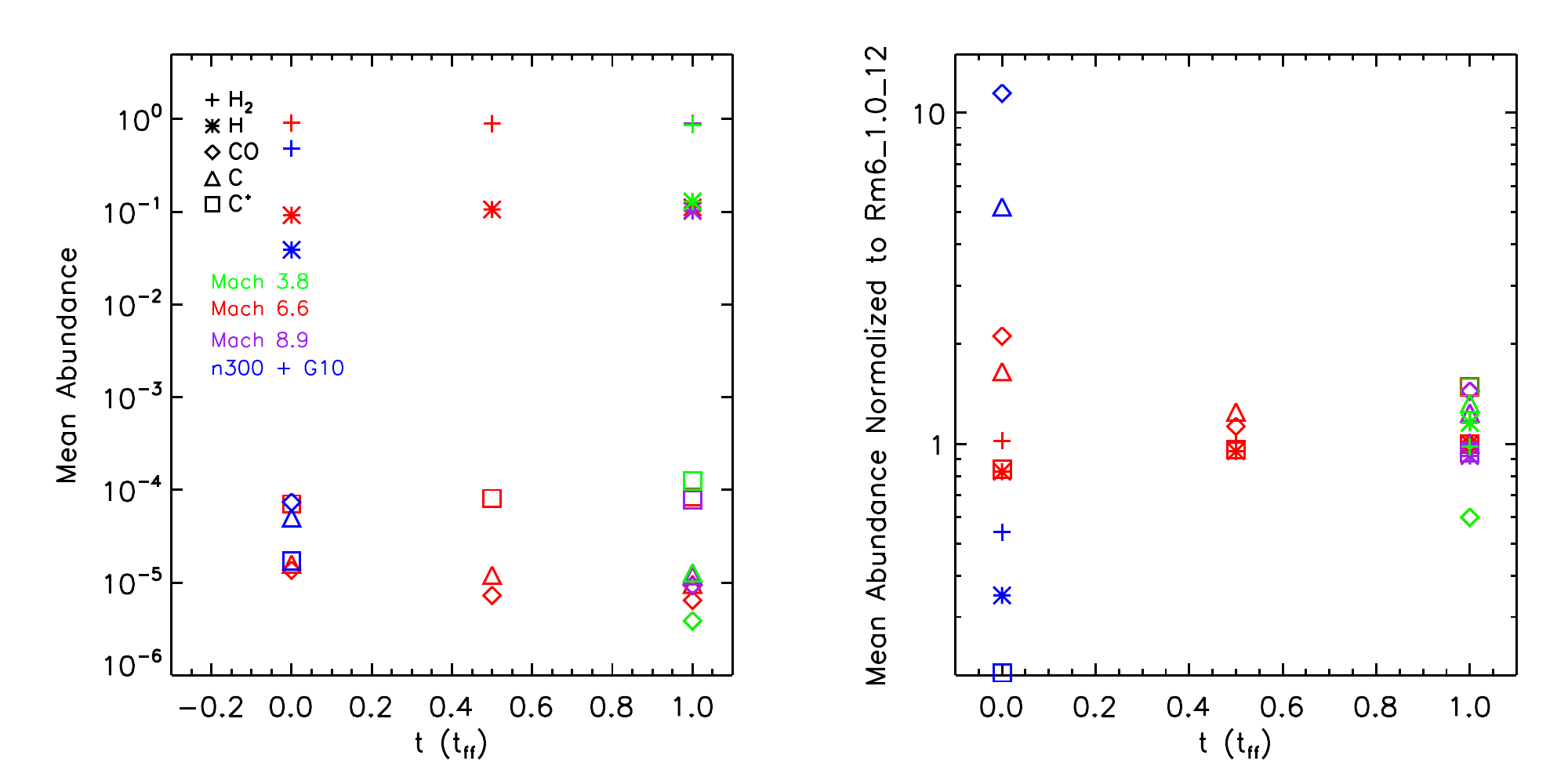}
\caption{Mean H, H$_2$, C$^+$, C, CO abundances for Rm6\_0.0\_12, Rm6\_0.5\_12, Rm6\_1.0\_12 (red), Rm9\_1.0\_12 (purple), Rm4\_1.0\_12 (green) and n300 using the G10 method (blue). On the right abundances have been normalized to the values of Rm6\_1.0\_12.
\label{mean_abund} }
\end{figure*}

Figure \ref{mean_abund} shows the mass-weighted abundances for simulations with two different Mach numbers at various evolutionary times. All {\sc orion} simulations have the same mean density such that apparent differences are directly due to variations in the gas morphology. 

Due to the relatively high simulation mean-density, we find that the H$_2$ abundance is fairly insensitive to changes in the density distribution caused by gravity. \citet{glover07a} found that the H$_2$ fraction {\it increased} with gravitational collapse for a smooth density distribution. However, in their case the initial mean densities were much lower than the initial density of our runs. Thus, in our simulations the gas is predominantly molecular at all times since most gas parcels are well-shielded with or without gravity.

We find that C and C$^+$ abundances vary by less than 20\%. The CO abundance demonstrates the largest variation and declines by more than a factor of two as the gas becomes self-gravitating. Much of this effect is due to a non-negligible fraction of the mass becoming concentrated in small, dense and well-shielded volumes that have fixed, maximal $10^{-4}$ abundance, which decreases the extinction in the remainder of the volume (see Figure \ref{comp_nhist}). Differences of $\sim$30\% in 3D Mach number have a relatively small impact (less than a factor of 2) on the total mean abundance but CO does show some sensitivity 
with abundance increasing with Mach number. H$_2$ and C$^+$ fractions decrease slightly with increasing Mach number.

For reference, Figure \ref{mean_abund} includes the mean abundances of n300. In order to compare the PDR results we only consider n300 gas with densities exceeding 200 cm$^{-3}$.\footnote{The n300 mean abundances are fairly similar whether or not the lower density gas is included because the averages are mass-weighted.} In this case, the abundance differences are dominated by the different mean extinctions. n300 has $\bar{Av} \sim 0.04$ while Rm6 has $\bar{Av} \sim 0.02$, which results in slightly lower mean C and CO abundances. The n300 mean H$_2$ abundance is lower than for those computed for Rm6; however, Figure 4 in G10 exhibits a higher H$_2$ abundance ($\sim 0.98$) for a simulation with the same mean density but lower magnetic field. This suggests that there is a morphological component to the H$_2$ abundance difference that is related to the magnetic field strength (S. Glover private communication).

\begin{figure*}
\epsscale{1.0}
\plotone{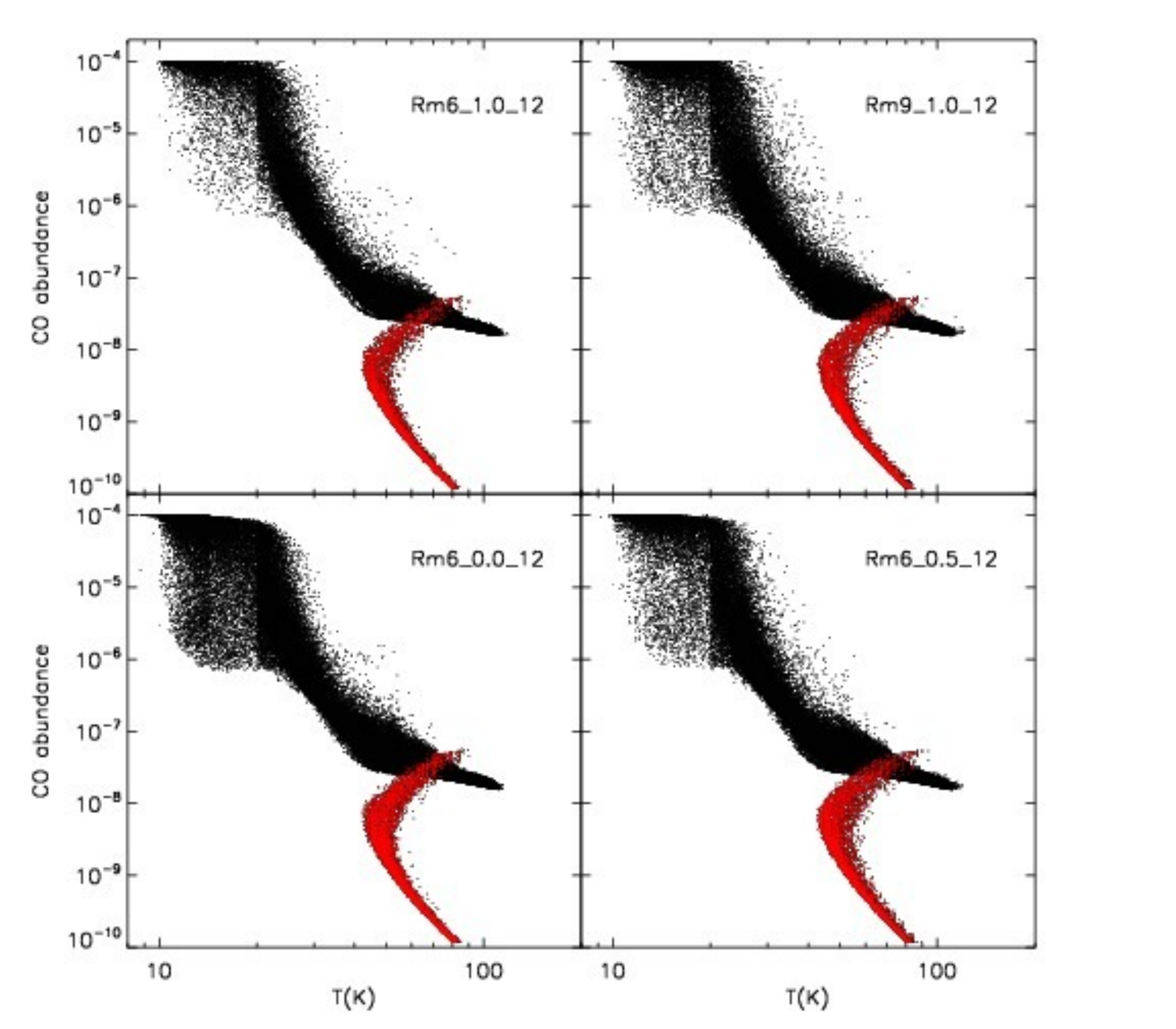}
\caption{CO abundance as a function of gas temperature for different times and Mach numbers. The cells that are within 2\% of the box boundary are colored red.
\label{temp_dist} }
\end{figure*}

Figure \ref{temp_dist} illustrates the CO distribution as a function of gas temperature. As gravity influences the gas distribution, the number of cells with high CO abundance and cold temperature ($10 \leq T \leq 20$) increases. This is related to the volume filling factor of the dense gas, which decreases as gas becomes more concentrated in dense, collapsing regions. The shape of the temperature-abundance distribution is otherwise roughly constant with Mach number and time. 

In all panels, points near the edge of the simulation box comprise a distinct swath of high-temperature/low-abundance points.  We color points within 2\% of the edge red to highlight this dichotomy. This region directly corresponds to the low-$A_{\rm v}$, mostly atomic region at the boundary. %We discuss the edge distribution in more detail in section \ref{edge}.

\subsection{Dependence on External Radiation Field}

\begin{figure}
\epsscale{1.0}
\plotone{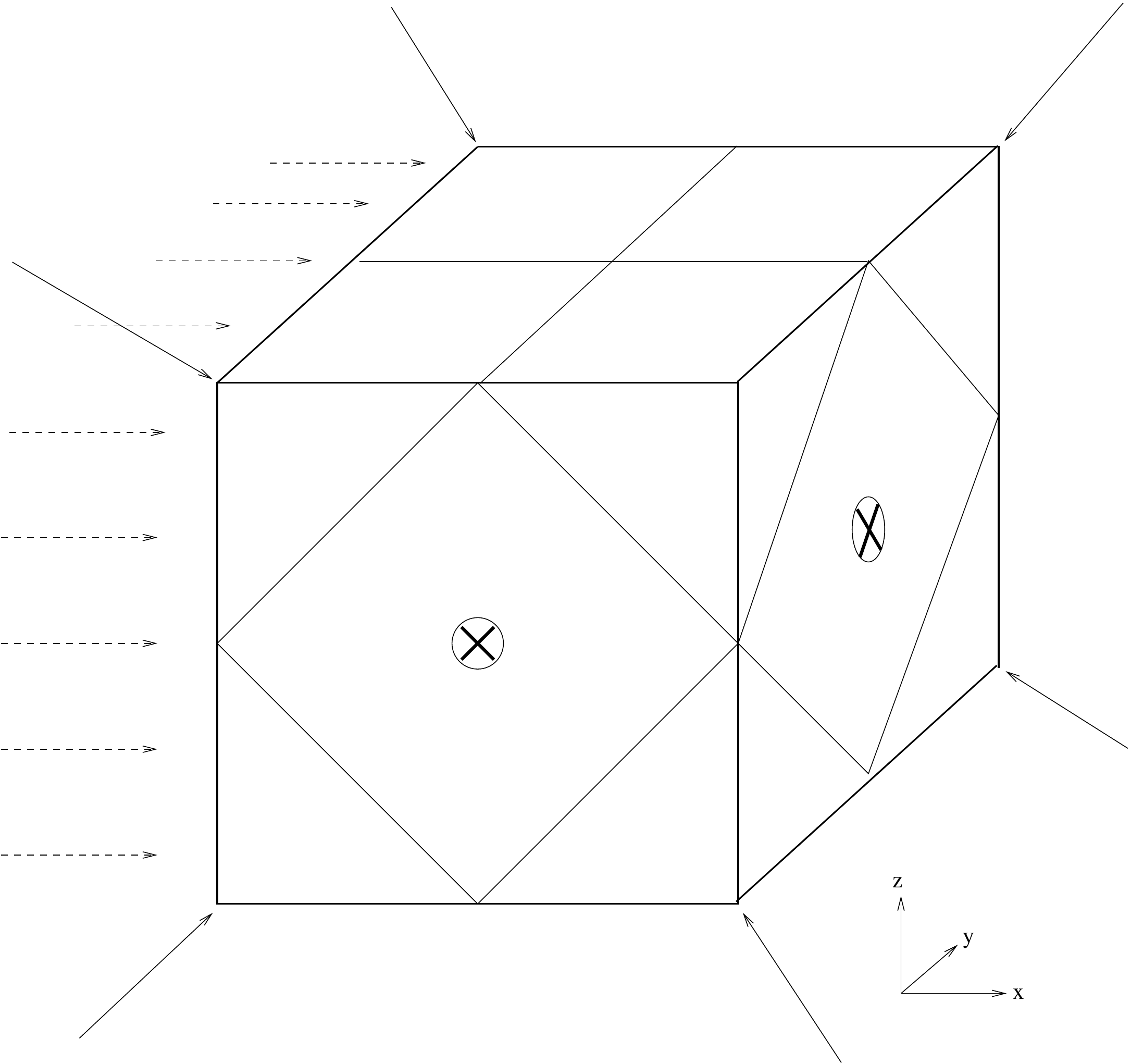}
\caption{Schematic of the incident radiation field used in the {\sc 3d-pdr} runs. The cube represents the entire computational domain. The thin solid lines represent the boundaries of the 12-ray {\sc healpix} structure as they are emanated from a point placed in the center of the computational domain. The direction of the isotropic radiation field is opposite to the direction of each {\sc healpix} ray as shown in the figure. The additional dashed lines on the left represent the direction of the plane-parallel radiation field added in run Rm6\_1.0\_12ui.
\label{cartoon} }
\end{figure}

In order to investigate how abundance depends on the external radiation field,  we consider two {\sc 3d-pdr} calculations with external fields each with a magnitude of 1 Draine but with different vectors. Run Rm6\_1.0\_12i has an external isotropic radiation field, while Run Rm6\_1.0\_12ui has an external field that is a superposition of a half Draine isotropic field and a half Draine uniform field (i.e., a field that is plane parallel to the simulation boundary at all faces). Figure \ref{cartoon} illustrates the incident radiation field geometry for the two cases.

\begin{figure}
\epsscale{1.25}
\plotone{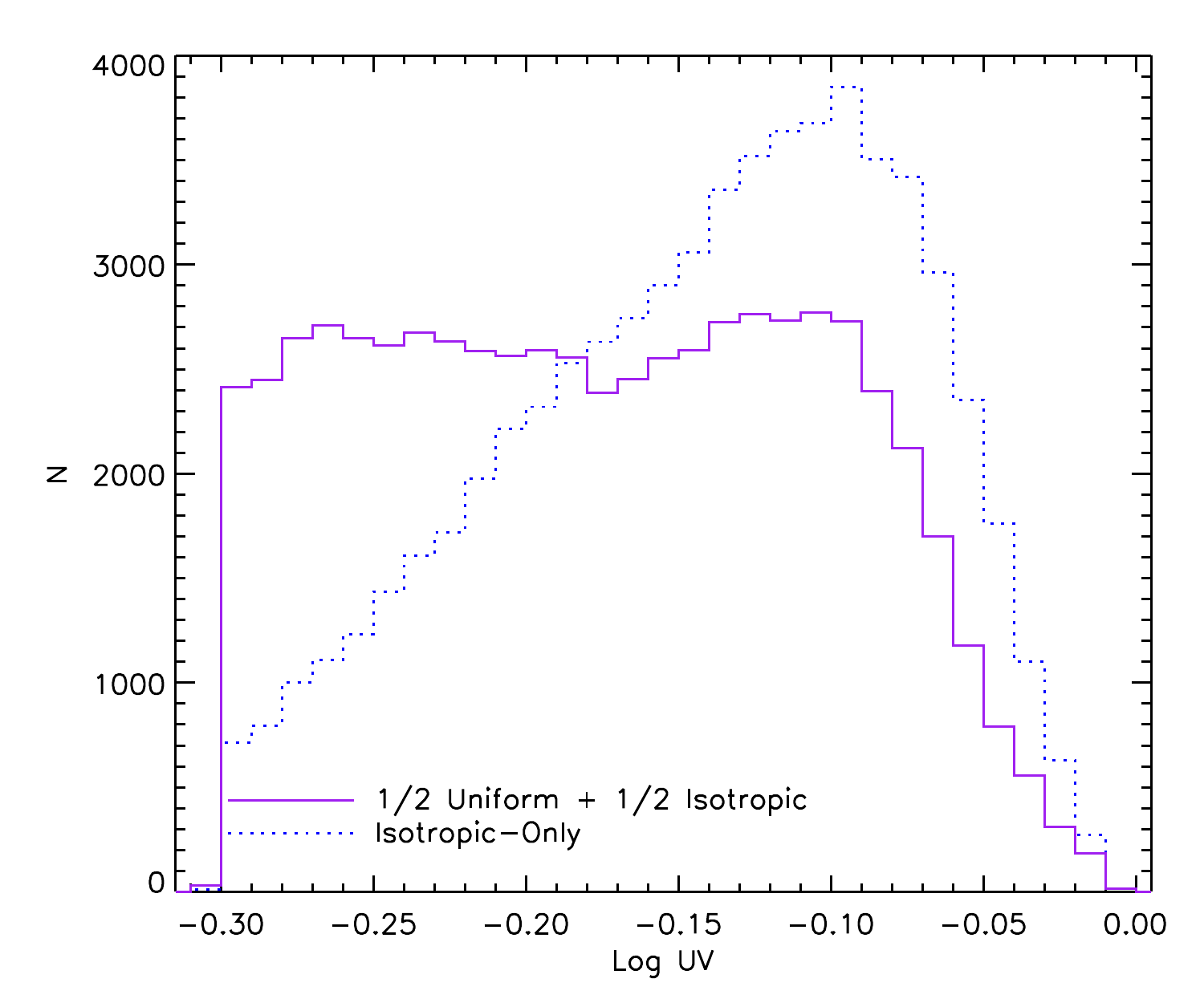}
\caption{Distribution of UV field for the Rm6\_1.0\_12i run which has an isotropic-only external 1 Draine field (blue, dotted) and the Rm6\_1.0\_12ui run which has a 0.5 Draine isotropic and 0.5 Draine uniform external field (solid, purple). Only the points that have a net field greater than 0.5 Draines are plotted. The two {\sc 3d-pdr} runs use the same input density distribution.
\label{uv_uni_iso} }
\end{figure}

\begin{figure*}
\epsscale{1.0}
\plotone{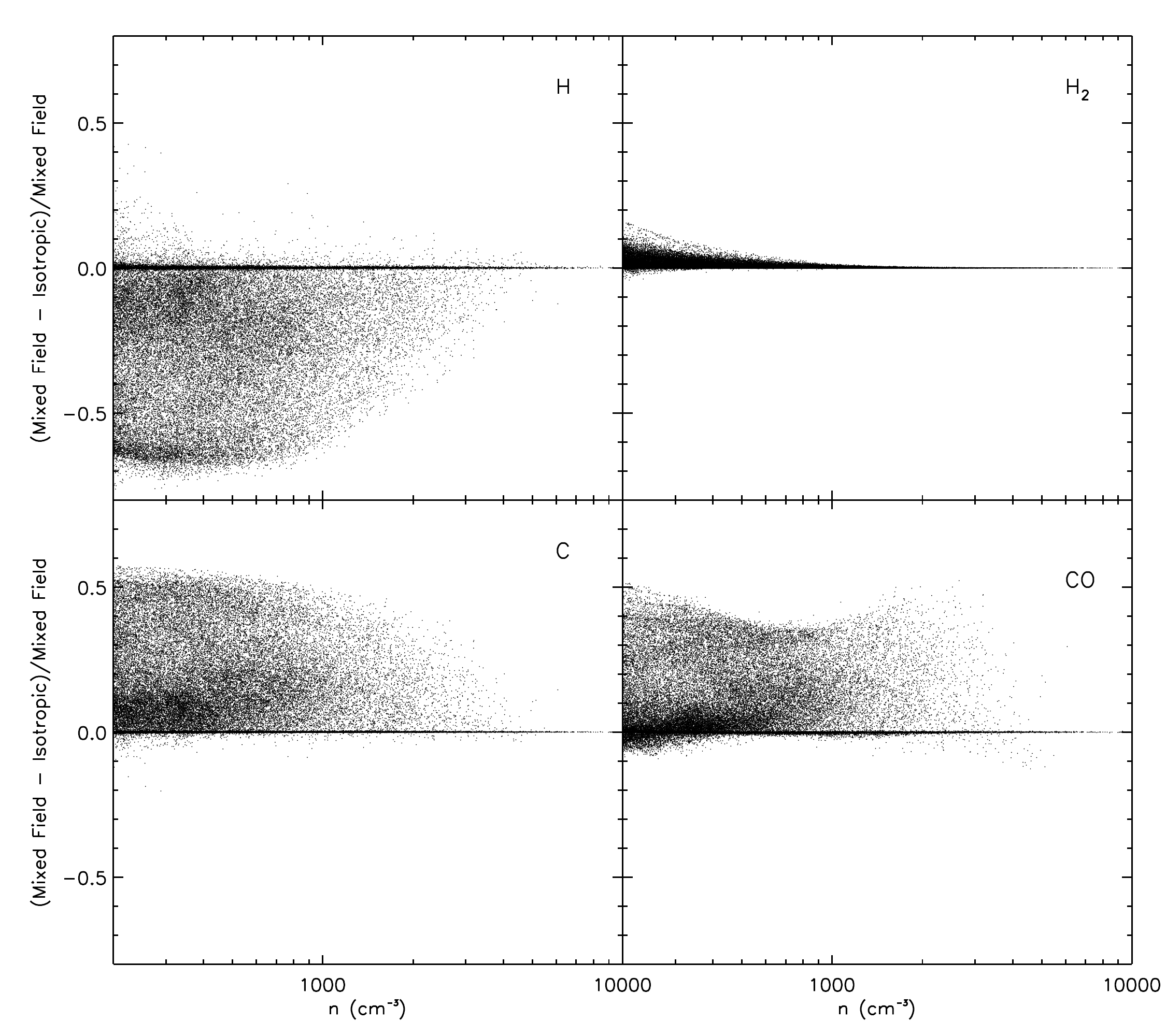}
\caption{Normalized relative differences between the H, H$_2$, C and CO abundances for Rm6\_1.0\_12i and Rm6\_1.0\_12ui (the same runs as in Figure \ref{uv_uni_iso}).
\label{diff_uni_iso} }
\end{figure*}

Figure \ref{uv_uni_iso} illustrates that by simply changing the field incidence the internal point-by-point UV distribution is very different.
%SSRO
The figure includes only points with a net field greater than 0.5 Draines; these points are ones which feel both the isotropic and plane-parallel components of the incident field and thus display the maximum difference.
Figure \ref{diff_uni_iso} shows the effect of the field differences on the H, H$_2$, C, and CO fractional abundances. Since CO abundance depends mainly on the local UV field and these distributions are distinct, it is unsurprising that individual abundances change by as much as $50\%$ for different field configurations. Likewise, C and H are strongly affected by the field distribution. Figure \ref{diff_uni_iso} shows that the mixed field simulation has fewer high UV points and more lower UV points, which is consistent with the elevated C and CO abundances displayed in Figure \ref{diff_uni_iso}. Since the molecular hydrogen abundance is nearly constant within the cloud, the field configuration at the boundary has little effect. The higher density gas ($n>10^3$ cm$^{-3}$), which is well self-shielded by definition, is also largely insensitive to field changes of this magnitude.

In summary, even a modest change in the UV field incidence reinforces the conclusion that three-dimensional PDR treatment is preferable to a one-dimensional treatment for complex or non-symmetric problems. We expect differences to be more significant for larger external field variations and for the inclusion of internal UV sources, i.e., protostars.
%We plan to explore the effects of including internal UV sources, i.e., protostars, in future work.

\section{Conclusions}\label{conclusions}

We use {\sc 3d-pdr} in combination with hydrodynamic molecular cloud simulations to explore the importance of dimensionality in PDR chemistry, to consider complex gas morphologies and to compare with prior results using an in situ astrochemistry treatment.

First, we demonstrate that our results are robust as a function of grid sampling and edge resolution. In fact, we find that the interior cloud abundances are remarkably insensitive to the resolution of the atomic to molecular transition at the cloud boundary. 

%Currently, both chemical post-processing and in situ chemistry approaches have limitations. Post-processing hydrodynamical simulations with {\sc 3d-pdr} provides significant time-savings and allows a larger chemical network in comparison to performing the same calculation in situ. 
We obtain reasonable agreement between the G10 in situ and {\sc 3d-pdr} approaches for the C and CO abundance distributions. This is because molecules such as CO and H$_2$ are not particularly sensitive to the dynamical history of the gas but instead depend predominantly on the local radiation field. 
%Thus, approaches that model the radiaiton field with good accuracy should produce similar results.
The two approaches differ the most for H and H$_2$ abundances near the cloud boundary and for cells that have low-extinction. For example, in G10 hydrogen is either entirely molecular or fully dissociated. This discrepancy appears to result from differences in the methodologies rather than chemical details, and we assert that the treatment of {\sc 3d-pdr} should be more accurate in transition regions.

We demonstrate that morphological differences due to cloud Mach number and evolutionary time can produce significant differences in the abundance distributions. While this may be difficult to observe directly since point by point abundances are difficult to infer, it may indirectly impact the  properties of the observed molecular emission lines emerging from the cloud.

Finally, we find that a relatively modest change in the external UV radiation field produces large changes in the chemical abundances. This supports the finding by B12 that three-dimensional treatment is crucial for complex and non-symmetric problems.

In paper II, we plan to implement several improvements to our method. First, we will relax our simple abundance approximations at densities $>10^4$ cm$^{-3}$ and instead couple {\sc 3d-pdr} to a one-zone gas-grain chemical network code. This will allow us to include molecular freezeout onto dust grains, gas-grain thermal coupling, and a more extensive chemical network. Second, we will investigate the effect of embedded UV sources on the chemical distribution. The {\sc orion} simulations that we analyze here contain detailed information about the masses and accretion rates of embedded protostars that we have neglected in this study. In addition to these changes, work is ongoing to couple {\sc 3d-pdr} to the photoionization and radiative transfer code {\sc MOCASSIN}.

\acknowledgements{The authors thank Simon Glover for helpful discussions and thank the referee, Robert Fisher, for suggestions that significantly improved the paper. The authors acknowledge support from NSF grant AST-0901055 (S.S.R.O), NASA grant HF-51311.01 (S.S.R.O), STFC grant ST/J001511/1 (T.G.B), and a JAE-DOC research contract (T.A.B.). T.A.B also thanks the Spanish MINECO for funding support through grants AYA2009-07304 and CSD200900038. The {\sc orion} simulations were performed on the Trestles XSEDE cluster.}

%\bibliography{clusterbib.bib}
%\bibliographystyle{apj}

\appendix
\section{Heating Rates }\label{heatingrate}

%Chemical heating is exothermic, which I guess is just H2 formation

\begin{figure}
\epsscale{0.8}
\plotone{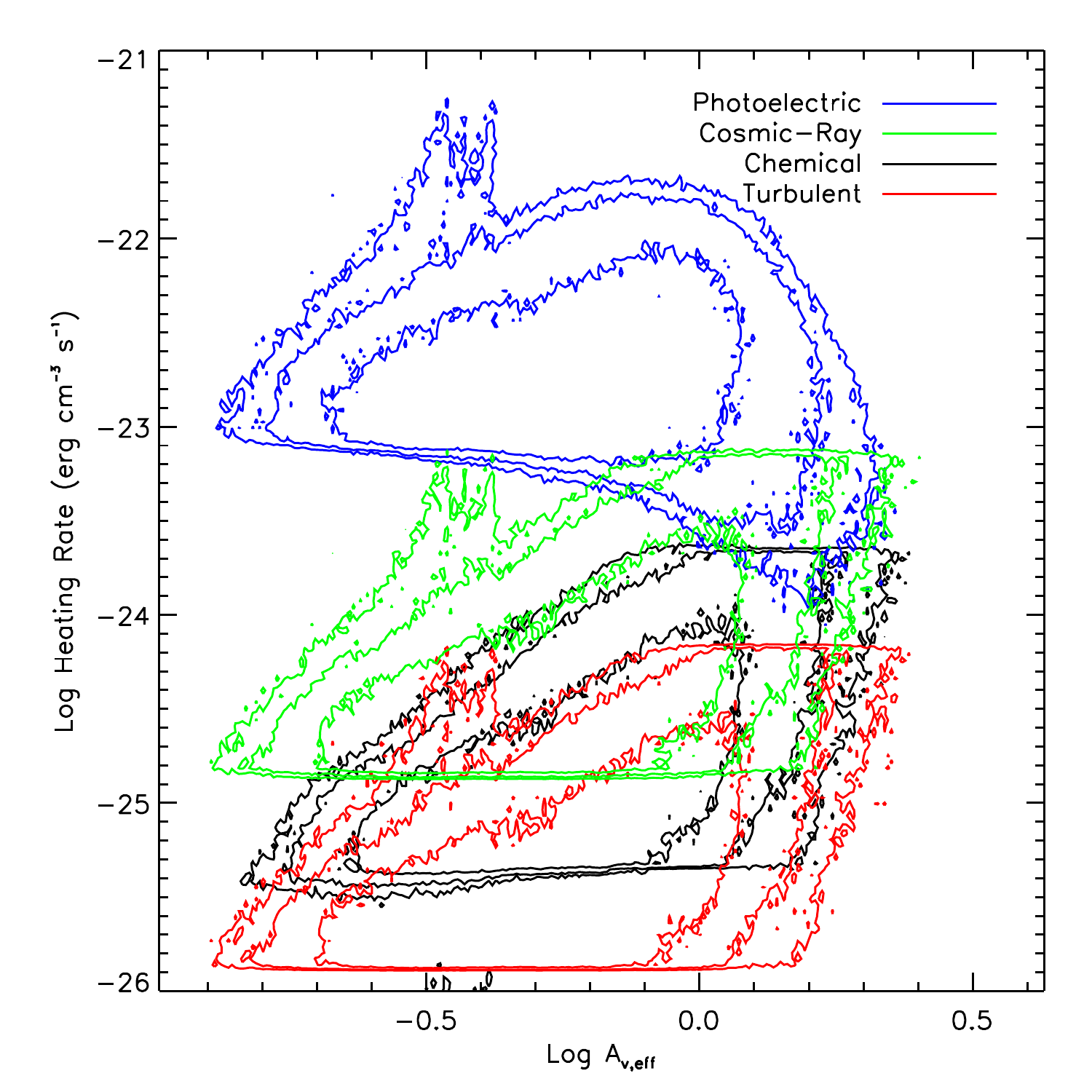}
\caption{Distribution of heating rates in Rm6\_1.0\_12 as a function of effective extinction for the total photoelectric (blue, top contours), cosmic-ray (green, top-middle contours), chemical (black, bottom-middle contours) and turbulent heating (red, bottom contours) in each cell. The contours correspond to the number of cells, $n$, with a given extinction and heating rate: $n=$100 (inner contour), $n=30 $ (middle contour), and $n=10$ (outer contour). 
\label{heating_dist} }
\end{figure}

There are four main contributions to the local heating at each domain point. First, there is photoelectric heating, which is produced by UV photon interactions with dust grains and PAHs, and which dominates near the cloud surface. Second, there is cosmic-ray ionization heating, which is set by the standard cosmic-ray density and becomes dominant deeper into the cloud. Third, there is chemical heating as the result of various exothermic reactions. Finally, there is turbulent heating, which is due to energy dissipation through shocks and depends on the turbulent outer scale (e.g., cloud size) and turbulent Mach number. Figure \ref{heating_dist} shows the distribution of heating rates for each of these contributions. Photoelectric heating dominates in most cases, and the turbulent heating generally provides the smallest contribution. If the turbulent heating is proportional to $v_{\rm TURB}^3/L$, then for the 2pc simulations here it should range from $v_{\rm TURB}^3/L=1.5\times 10^{-5}-1.4\times 10^{-4}$ cm$^2$ s$^{-3}$, which brackets the constant value, $6.5 \times 10^{-5}$  cm$^2$ s$^{-3}$, we adopt in our calculations. We direct the reader to \citet{pan09} and \citet{kumar13} for additional discussion and modeling of heating due to turbulent dissipation, intermittency, and shear flows. 
% Note using (1D Mach *cs)^3/2pc for the fiducial run gives 6*10^-5, within 10%

\bibliography{turb_pdr_pdf.bib}

\begin{thebibliography}{46}
\expandafter\ifx\csname natexlab\endcsname\relax\def\natexlab#1{#1}\fi

\bibitem[{{Banerjee} {et~al.}(2009){Banerjee}, {V{\'a}zquez-Semadeni},
  {Hennebelle}, \& {Klessen}}]{banerjee09}
{Banerjee}, R., {V{\'a}zquez-Semadeni}, E., {Hennebelle}, P., \& {Klessen},
  R.~S. 2009, \mnras, 398, 1082

\bibitem[{{Bayet} {et~al.}(2010){Bayet}, {Hartquist}, {Viti}, {Williams}, \&
  {Bell}}]{bayet10}
{Bayet}, E., {Hartquist}, T.~W., {Viti}, S., {Williams}, D.~A., \& {Bell},
  T.~A. 2010, \aap, 521, A16

\bibitem[{{Bayet} {et~al.}(2009){Bayet}, {Viti}, {Williams}, {Rawlings}, \&
  {Bell}}]{bayet09}
{Bayet}, E., {Viti}, S., {Williams}, D.~A., {Rawlings}, J.~M.~C., \& {Bell}, T.
  2009, \apj, 696, 1466

\bibitem[{{Bell} {et~al.}(2006){Bell}, {Roueff}, {Viti}, \&
  {Williams}}]{bell06}
{Bell}, T.~A., {Roueff}, E., {Viti}, S., \& {Williams}, D.~A. 2006, \mnras,
  371, 1865

\bibitem[{{Bergin} {et~al.}(2004){Bergin}, {Hartmann}, {Raymond}, \&
  {Ballesteros-Paredes}}]{bergin04}
{Bergin}, E.~A., {Hartmann}, L.~W., {Raymond}, J.~C., \& {Ballesteros-Paredes},
  J. 2004, \apj, 612, 921

\bibitem[{{Bisbas} {et~al.}(2012){Bisbas}, {Bell}, {Viti}, {Yates}, \&
  {Barlow}}]{bisbas12}
{Bisbas}, T.~G., {Bell}, T.~A., {Viti}, S., {Yates}, J., \& {Barlow}, M.~J.
  2012, \mnras, 427, 2100

\bibitem[{{Castor}(1970)}]{castor70}
{Castor}, J.~I. 1970, \mnras, 149, 111

\bibitem[{{de Jong} {et~al.}(1975){de Jong}, {Dalgarno}, \& {Chu}}]{dejong75}
{de Jong}, T., {Dalgarno}, A., \& {Chu}, S.-I. 1975, \apj, 199, 69

\bibitem[{{Ercolano} {et~al.}(2005){Ercolano}, {Barlow}, \&
  {Storey}}]{ercolano05}
{Ercolano}, B., {Barlow}, M.~J., \& {Storey}, P.~J. 2005, \mnras, 362, 1038

\bibitem[{{Ercolano} {et~al.}(2003){Ercolano}, {Barlow}, {Storey}, \&
  {Liu}}]{ercolano03}
{Ercolano}, B., {Barlow}, M.~J., {Storey}, P.~J., \& {Liu}, X.-W. 2003, \mnras,
  340, 1136

\bibitem[{{Ercolano} {et~al.}(2008){Ercolano}, {Young}, {Drake}, \&
  {Raymond}}]{ercolano08}
{Ercolano}, B., {Young}, P.~R., {Drake}, J.~J., \& {Raymond}, J.~C. 2008,
  \apjs, 175, 534

\bibitem[{{Glover} \& {Clark}(2012)}]{glover12}
{Glover}, S.~C.~O. \& {Clark}, P.~C. 2012, \mnras, 426, 377

\bibitem[{{Glover} {et~al.}(2010){Glover}, {Federrath}, {Mac Low}, \&
  {Klessen}}]{glover10}
{Glover}, S.~C.~O., {Federrath}, C., {Mac Low}, M.-M., \& {Klessen}, R.~S.
  2010, \mnras, 404, 2

\bibitem[{{Glover} \& {Mac Low}(2007{\natexlab{a}})}]{glover07a}
{Glover}, S.~C.~O. \& {Mac Low}, M.-M. 2007{\natexlab{a}}, \apjs, 169, 239

\bibitem[{{Glover} \& {Mac Low}(2007{\natexlab{b}})}]{glover07b}
---. 2007{\natexlab{b}}, \apj, 659, 1317

\bibitem[{{Glover} \& {Mac Low}(2011)}]{glover11}
---. 2011, \mnras, 412, 337

%\bibitem[{{G{\'o}rski} {et~al.}(2005){G{\'o}rski}, {Hivon}, {Banday},
%  {Wandelt}, {Hansen}, {Reinecke}, \& {Bartelmann}}]{gorski05}
%{G{\'o}rski}, K.~M., {Hivon}, E., {Banday}, A.~J., {Wandelt}, B.~D., {Hansen},
%  F.~K., {Reinecke}, M., \& {Bartelmann}, M. 2005, \apj, 622, 759

\bibitem[{{G{\'o}rski} {et~al.}(2005){G{\'o}rski}, {Hivon}, {Banday},
  {Wandelt}, {Hansen}, {Reinecke}, \& {Bartelmann}}]{gorski05}
{G{\'o}rski}, K.~M., {Hivon}, E., {Banday}, A.~J., et al. 2005, \apj, 622, 759

\bibitem[{{Hansen} {et~al.}(2012){Hansen}, {Klein}, {McKee}, \&
  {Fisher}}]{hansen12}
{Hansen}, C.~E., {Klein}, R.~I., {McKee}, C.~F., \& {Fisher}, R.~T. 2012, \apj,
  747, 22

\bibitem[{{Hollenbach} {et~al.}(1971){Hollenbach}, {Werner}, \&
  {Salpeter}}]{hollenbach71}
{Hollenbach}, D.~J., {Werner}, M.~W., \& {Salpeter}, E.~E. 1971, \apj, 163, 165

\bibitem[{{Klein}(1999)}]{klein99}
{Klein}, R.~I. 1999, JCoAM, 109, 123

\bibitem[{{Kritsuk} {et~al.}(2007){Kritsuk}, {Norman}, {Padoan}, \&
  {Wagner}}]{kritsuk07}
{Kritsuk}, A.~G., {Norman}, M.~L., {Padoan}, P., \& {Wagner}, R. 2007, \apj,
  665, 416

\bibitem[{{Krumholz} {et~al.}(2007){Krumholz}, {Klein}, \&
  {McKee}}]{krumholz07}
{Krumholz}, M.~R., {Klein}, R.~I., \& {McKee}, C.~F. 2007, \apj, 656, 959

\bibitem[{{Krumholz} {et~al.}(2004){Krumholz}, {McKee}, \&
  {Klein}}]{krumholz04}
{Krumholz}, M.~R., {McKee}, C.~F., \& {Klein}, R.~I. 2004, \apj, 611, 399

\bibitem[{{Kumar} \& {Fisher}(2013)}]{kumar13}
{Kumar}, A. \& {Fisher}, R.~T. 2013, \mnras, 431, 455 

\bibitem[{{Le Teuff} {et~al.}(2000){Le Teuff}, {Millar}, \&
  {Markwick}}]{leteuff00}
{Le Teuff}, Y.~H., {Millar}, T.~J., \& {Markwick}, A.~J. 2000, \aaps, 146, 157

%\bibitem[{{Levrier} {et~al.}(2012){Levrier}, {Le Petit}, {Hennebelle},
%  {Lesaffre}, {Gerin}, \& {Falgarone}}]{levrier12}
%{Levrier}, F., {Le Petit}, F., {Hennebelle}, P., {Lesaffre}, P., {Gerin}, M.,
%  \& {Falgarone}, E. 2012, \aap, 544, A22

\bibitem[{{Levrier} {et~al.}(2012){Levrier}, {Le Petit}, {Hennebelle},
  {Lesaffre}, {Gerin}, \& {Falgarone}}]{levrier12}
{Levrier}, F., {Le Petit}, F., {Hennebelle}, P., et al. 2012, \aap, 544, A22

\bibitem[{{Mac Low}(1999)}]{maclow99}
{Mac Low}, M.-M. 1999, \apj, 524, 169

\bibitem[{{Mac Low} \& {Klessen}(2004)}]{maclow04}
{Mac Low}, M.-M. \& {Klessen}, R.~S. 2004, RvMP, 76, 125

\bibitem[{{Masunaga} {et~al.}(1998){Masunaga}, {Miyama}, \&
  {Inutsuka}}]{masunaga98}
{Masunaga}, H., {Miyama}, S.~M., \& {Inutsuka}, S.-I. 1998, \apj, 495, 346

\bibitem[{{McKee} \& {Ostriker}(2007)}]{mckee07}
{McKee}, C.~F. \& {Ostriker}, E.~C. 2007, \araa, 45, 565

\bibitem[{{Nelson} \& {Langer}(1997)}]{nelson97}
{Nelson}, R.~P. \& {Langer}, W.~D. 1997, \apj, 482, 796

\bibitem[{{Nelson} \& {Langer}(1999)}]{nelson99}
---. 1999, \apj, 524, 923

\bibitem[{{Offner} {et~al.}(2008){Offner}, {Klein}, \& {McKee}}]{Offner08a}
{Offner}, S. S.~R., {Klein}, R.~I., \& {McKee}, C.~F. 2008, \apj, 681, 375

\bibitem[{{Offner} {et~al.}(2009){Offner}, {Klein}, {McKee}, \&
  {Krumholz}}]{Offner09a}
{Offner}, S.~S.~R., {Klein}, R.~I., {McKee}, C.~F., \& {Krumholz}, M.~R. 2009,
  \apj, 703, 131

%\bibitem[{{Osterbrock}(1974)}]{osterbrock74}
%{Osterbrock}, D.~E. 1974, {Astrophysics of gaseous nebulae} (Freeman \& Co.,
%  San Francisco)

\bibitem[Osterbrock(1974)]{osterbrock74} Osterbrock, D.~E.\ 1974, 
Research supported by the Research Corp., Wisconsin Alumni Research 
Foundation, John Simon Guggenheim Memorial Foundation, Institute for 
Advanced Studies, and National Science Foundation.~San Francisco, 
W.~H.~Freeman and Co., 1974.~263 p.  

\bibitem[{{Padoan} {et~al.}(1997){Padoan}, {Nordlund}, \& {Jones}}]{padoan97}
{Padoan}, P., {Nordlund}, A., \& {Jones}, B.~J.~T. 1997, \mnras, 288, 145

\bibitem[{{Pan} \& {Padoan}(2009)}]{pan09}
{Pan}, L. \& {Padoan}, P. 2009, \apj, 692, 594

\bibitem[{{Pavlovski} {et~al.}(2006){Pavlovski}, {Smith}, \& {Mac
  Low}}]{pavlovski06}
{Pavlovski}, G., {Smith}, M.~D., \& {Mac Low}, M.-M. 2006, \mnras, 368, 943

\bibitem[{{Pavlovski} {et~al.}(2002){Pavlovski}, {Smith}, {Mac Low}, \&
  {Rosen}}]{pavlovski02}
{Pavlovski}, G., {Smith}, M.~D., {Mac Low}, M.-M., \& {Rosen}, A. 2002, \mnras,
  337, 477

%\bibitem[{{R{\"o}llig} {et~al.}(2007){R{\"o}llig}, {Abel}, {Bell}, {Bensch},
%  {Black}, {Ferland}, {Jonkheid}, {Kamp}, {Kaufman}, {Le Bourlot}, {Le Petit},
%  {Meijerink}, {Morata}, {Ossenkopf}, {Roueff}, {Shaw}, {Spaans}, {Sternberg},
%  {Stutzki}, {Thi}, {van Dishoeck}, {van Hoof}, {Viti}, \&
%  {Wolfire}}]{rollig07}
%{R{\"o}llig}, M., {Abel}, N.~P., {Bell}, T., {Bensch}, F., {Black}, J.,
%  {Ferland}, G.~J., {Jonkheid}, B., {Kamp}, I., {Kaufman}, M.~J., {Le Bourlot},
%  J., {Le Petit}, F., {Meijerink}, R., {Morata}, O., {Ossenkopf}, V., {Roueff},
%  E., {Shaw}, G., {Spaans}, M., {Sternberg}, A., {Stutzki}, J., {Thi}, W.-F.,
%  {van Dishoeck}, E.~F., {van Hoof}, P.~A.~M., {Viti}, S., \& {Wolfire}, M.~G.
%  2007, \aap, 467, 187

\bibitem[{{R{\"o}llig} {et~al.}(2007){R{\"o}llig}, {Abel}, {Bell}, {Bensch},
  {Black}, {Ferland}, {Jonkheid}, {Kamp}, {Kaufman}, {Le Bourlot}, {Le Petit},
  {Meijerink}, {Morata}, {Ossenkopf}, {Roueff}, {Shaw}, {Spaans}, {Sternberg},
  {Stutzki}, {Thi}, {van Dishoeck}, {van Hoof}, {Viti}, \&
  {Wolfire}}]{rollig07}
{R{\"o}llig}, M., {Abel}, N.~P., {Bell}, T., et al. 2007, \aap, 467, 187

\bibitem[{{Shetty} {et~al.}(2011){Shetty}, {Glover}, {Dullemond}, \&
  {Klessen}}]{shetty11}
{Shetty}, R., {Glover}, S.~C., {Dullemond}, C.~P., \& {Klessen}, R.~S. 2011,
  \mnras, 412, 1686

%\bibitem[{{Sobolev}(1960)}]{sobolev60}
%{Sobolev}, V.~V. 1960, {Moving envelopes of stars} (Harvard Univ. Press,
%  Cambridge)

\bibitem[Sobolev(1960)]{sobolev60} Sobolev, V.~V.\ 1960, 
Cambridge: Harvard University Press, 1960  

%\bibitem[{{Truelove} {et~al.}(1997){Truelove}, {Klein}, {McKee}, {Holliman},
%  {Howell}, \& {Greenough}}]{truelove97}
%{Truelove}, J.~K., {Klein}, R.~I., {McKee}, C.~F., {Holliman}, II, J.~H.,
%  {Howell}, L.~H., \& {Greenough}, J.~A. 1997, \apjl, 489, L179+

\bibitem[{{Truelove} {et~al.}(1997){Truelove}, {Klein}, {McKee}, {Holliman},
  {Howell}, \& {Greenough}}]{truelove97}
{Truelove}, J.~K., {Klein}, R.~I., {McKee}, C.~F., et al. 1997, \apjl, 489, L179+

%\bibitem[{{Truelove} {et~al.}(1998){Truelove}, {Klein}, {McKee}, {Holliman},
%  {Howell}, {Greenough}, \& {Woods}}]{truelove98}
%{Truelove}, J.~K., {Klein}, R.~I., {McKee}, C.~F., {Holliman}, II, J.~H.,
%  {Howell}, L.~H., {Greenough}, J.~A., \& {Woods}, D.~T. 1998, \apj, 495, 821

\bibitem[{{Truelove} {et~al.}(1998){Truelove}, {Klein}, {McKee}, {Holliman},
  {Howell}, {Greenough}, \& {Woods}}]{truelove98}
{Truelove}, J.~K., {Klein}, R.~I., {McKee}, C.~F., et al. 1998, \apj, 495, 821

\bibitem[{{Van Loo} {et~al.}(2013){Van Loo}, {Butler}, \& {Tan}}]{vanloo13}
{Van Loo}, S., {Butler}, M.~J., \& {Tan}, J.~C. 2013, \apj, 764, 36

\bibitem[{{Woodall} {et~al.}(2007){Woodall}, {Ag{\'u}ndez}, {Markwick-Kemper},
  \& {Millar}}]{woodall07}
{Woodall}, J., {Ag{\'u}ndez}, M., {Markwick-Kemper}, A.~J., \& {Millar}, T.~J.
  2007, \aap, 466, 1197

\end{thebibliography}
\bibliographystyle{apj}

\end{document}